\documentclass[a4paper,11pt]{article}

\pdfoutput=1

\usepackage{a4wide}
\usepackage{graphicx}
\usepackage[usenames]{color}
\usepackage{amssymb}

\usepackage{amsmath}
\usepackage{verbatim}
\usepackage{units}

\usepackage{cases}

\usepackage[most]{tcolorbox}
\usepackage{doi}
\usepackage{hyperref}

\begin{document}

\begin{center}   
\textbf{\LARGE Modeling the dynamics of aeolian meter-scale bedforms induced by bed heterogeneities}

\vspace*{0.2cm}

C. \textsc{Rambert}$^a$,
J.M. \textsc{Nield}$^b$,
C. \textsc{Narteau}$^c$,
P. \textsc{Delorme}$^d$,
G.F.S. \textsc{Wiggs}$^e$,\\
M.C. \textsc{Baddock}$^f$,
J. \textsc{Best}$^g$,
K.T. \textsc{Christensen}$^h$
and P. \textsc{Claudin}$^a$
\end{center}

{
\renewcommand{\baselinestretch}{0.9}
\footnotesize
\noindent
$^a$ {Physique et M\'ecanique des Milieux H\'et\'erog\`enes, CNRS, ESPCI Paris, PSL Research University, Universit\'e Paris Cit\'e, Sorbonne Universit\'e, Paris, France}\\
$^b$ {School of Geography and Environmental Science, University of Southampton, Southampton, UK}\\
$^c$ {Institut de Physique du Globe de Paris, Universit\'e Paris Cit\'e, CNRS, Paris, France}\\
$^d$ {Laboratoire de G\'eologie, Ecole Normale Sup\'erieure, CNRS, PSL Research University, Paris, France}\\
$^e$ {School of Geography and the Environment, University of Oxford, Oxford, UK}\\
$^f$ {Geography and Environment, Loughborough University, Loughborough, UK}\\
$^g$ {Departments of Earth Science and Environmental Change, Geography \& GIS, and Mechanical Science \& Engineering, and Ven Te Chow Hydrosystems Laboratory, University of Illinois at Urbana-Champaign, Urbana, IL 61801, USA}\\
$^h$ {Department of Mechanical Engineering, University of Colorado Denver, Denver, CO 80204, USA}
\par
}

\begin{abstract}
Desert surfaces are typically non uniform, with individual sand dunes generally surrounded by gravel or non-erodible beds. Similarly, beaches vary in composition and moisture that enhances cohesion between the grains. These bed heterogeneities affect the aeolian transport properties greatly, and can then influence the emergence and dynamics of bedforms. Here, we propose a model that describes how, due to transport capacity being greater on consolidated than erodible beds, patches of sand can grow, migrate and spread to form bedforms with meter-scale length. Our approach has a quantitative agreement with high-resolution spatio-temporal observations, where conventional theory would predict the disappearance of these small bedforms. A crucial component of the model is that the transport capacity does not instantly change from one bed configuration to another. Instead, transport capacity develops over a certain distance, which thereby determines the short-term evolution of the bedform. The model predicts various stages in the development of these meter-scale bedforms, and explains how the evolution of bed elevation profiles observed in the field depends on the duration of the wind event and the intensity of the incoming sand flux. Our study thus sheds light on the initiation and dynamics of early-stage bedforms by establishing links between surface properties, emerging sand patterns and protodunes, commonly observed in coastal and desert landscapes.
\end{abstract}

\begin{tcolorbox}[colback=blue!5!white,colframe=blue!75!black]
{
\renewcommand{\baselinestretch}{0.9}
\small
\centerline{\textbf{Significance}}
\vspace*{0.2cm}
We present a model to explain the emergence of meter-scale bedforms that grow in coarse-grained interdune areas or on moist beaches. We show that the existing theory of dune dynamics must be extended to account for the spatial variation of wind transport capacity over bed heterogeneities, with enhanced transport over consolidated rather than erodible surfaces. The quantitative agreement between the model predictions and a unique set of high-precision field data acquired in the Namib Desert allows us to theoretically explore the different dynamics of such emerging bedforms, which can eventually disappear or lead to dune formation. This work provides new ways to interpret the initiation and evolution of small bedforms, and facilitates the estimation of aeolian transport in diverse environments.
\par
}
\end{tcolorbox}

\begin{center}
Proc. Natl. Acad. Sci. USA \textbf{122}, e2426143122 (2025).\\
\href{https://doi.org/10.1073/pnas.2426143122}{\texttt{doi.org/10.1073/pnas.2426143122}}
\end{center}

\section{Introduction}
\label{intro}
Aeolian dunes are often sufficiently large to accommodate seasonal variations of surface winds, resulting in a diversity of shapes, orientations and dynamics according to sand availability and environmental conditions \cite{Cour24}. At the same time, smaller bedforms, either ephemeral or destined to grow into protodunes \cite{Kocu92,Mont20}, can emerge during short individual wind events \cite{Lanc96,Hesp97,Elbe05,Niel11a,Elbe12}. Such early stage bedforms are important to study as they reveal some of the key processes of dune dynamics at work before they become affected by more complicated phenomena such as coarsening and nonlinear effects \cite{Bris22}. Although some of these emerging bedforms have been clearly identified as resulting from the unstable nature of a flat sand bed with a length selected by the interplay between topography, wind flow and sediment transport \cite{Gada20a,Delo20,Lu21}, others, with sizes typically smaller than the cut-off length ($\simeq 10$~m) below which dunes are not expected to grow, exhibit their own morphodynamics \cite{Badd18}. This is particularly the case for small, meter-scale, sandy bedforms that develop over more consolidated beds. They have been associated with sediment transport properties that vary due to spatial heterogeneities of the substrate, including transitions between consolidated and erodible beds \cite{Delo23}. These bedforms are distinct from decimeter-scale wind ripples, which adorn the flanks of dunes and sandy surfaces in general, and whose dynamics is due to the properties of saltating particles when they impact a fully erodible granular bed \cite{Ande87,Csah00,Yizh04,Dura14,Lest25}. The manner in which such meter-scale bedforms may, or may not, grow into protodunes and dunes has yet to be investigated precisely, and this is one of the purposes of the present work.

Since the pioneering studies of Bagnold on the physics of wind-blown sand \cite{Bagn37,Bagn41}, the properties of aeolian saltation and, in particular, the quantification of the grain trajectories and transport capacity of a wind of a given strength have been investigated \cite{Nalp93,Rasm96,Iver99,Dong04,Rasm08,Crey09,Ho12,Ho14,Mart17} and reviewed \cite{Vala15,Paht20}. Saltation has also been shown to be sensitive to the nature of the bed over which the grains rebound, with enhanced transport over a consolidated surface (either solid, gravel or moist/cohesive) in comparison to that on an erodible sandy layer \cite{Ho11,Niel11b,OBri23}. Rebounds are less dissipative over a rigid surface than a granular one, resulting in higher saltation layers \cite{Ho11}. This has consequences for the feedback that moving grains exert on the wind flow. For ordinary saltation on sand beds, the transport layer of almost constant height (Bagnold's focal height) accommodates saltating grains at a large enough volume fraction to reduce the basal wind shear velocity to its transport threshold value \cite{Dura11,Vala15}. The velocity of the moving grains is thus constant, independent of wind speed \cite{Dura12}. By contrast, over consolidated beds, the grains, which move in a more expanded transport layer, have a weaker and more dilute feedback on the flow, so that their velocity typically follows that of the wind. This explains a ratio in transport capacities between consolidated and erodible beds that scales with the wind shear velocity \cite{Ho11,Jenk14,Berz16}.

As proposed by \cite{Niel11,Delo23}, reduction in sediment transport capacity where the bed transitions from consolidated to erodible is a way to amplify local sand deposition and trigger the development of small bedforms. The aim herein is to investigate more precisely by which processes, and at which scales, the growth and propagation of early-stage bedforms can occur. This also raises the question as to how the standard dune model, which predicts steady propagative flat domes at a size close to the cut-off length -- but not smaller -- almost independently of the incoming sand flux \cite{Andr02a}, can be generalized to describe these meter-scale growing bedforms, which are sensitive to sand supply. Here, we use field measurements of the evolution of small bedforms in the Namib Desert with a high spatial precision ($0.01$~m) and over a short time period ($\simeq 1$~hour). We test and develop components of a dune model in order to generalize the analysis of aeolian bedforms over a previously inaccessible range of length scales and through a transient dynamics that have not yet been explored.

The paper starts with the description of the field location and data acquisition. We then present the model, highlighting its additional components with respect to the reference dune emergence theory, especially a length scale that quantifies the distance over which the consolidated/erodible bed transition occurs. Depending on the parameter values, in particular the input sand flux and this length scale, different dynamics are found, by which an initially flat patch after a certain time either disappears, grows while migrating, or spreads both up and downwind. The fit of these parameters to reproduce quantitatively the spatio-temporal variations of the bedform profile measured in the field gives good confidence on the relevance of these processes associated with bed heterogeneities. Understanding of the microscopic nature of this transition length scale is, however, an essential question that remains to be investigated further. These results also open discussion on the longer-term evolution of such out-of-equilibrium bedforms, which can quickly appear, grow and/or disappear, and motivates future studies on their interactions and dynamics.

\begin{figure}[t!]
\centering
\includegraphics[width=\linewidth]{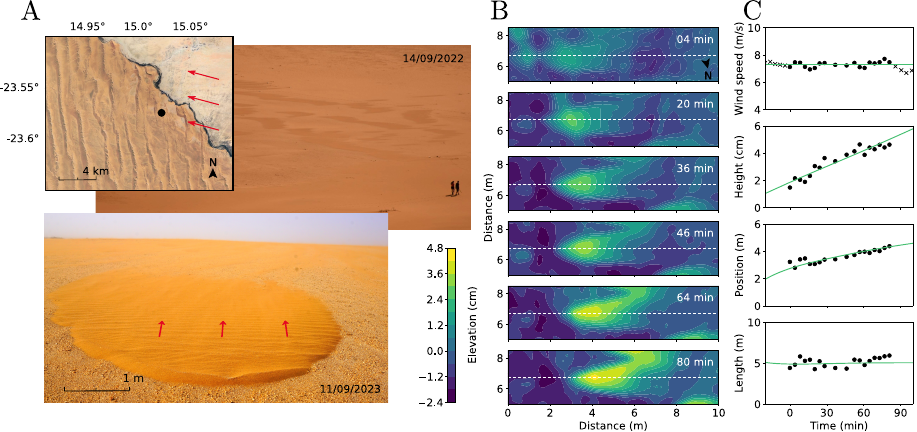}
\caption{
{\bf Morphodynamics of meter-scale sand bedforms at the northern border of the Namib Sand Sea, Namibia.}
({\bf A}) Location of the site close to Gobabeb Namib Research Institute (15$^\circ$01'31''E, 23$^\circ$34''31'S). Top photo: Aerial view of the site. Middle photo: View of the field of bedforms. Bottom photo: closer view of an individual bedform. Red arrows show wind direction.
({\bf B}) 
Elevation maps of a growing and migrating bedform measured over 80 minutes (color bar on the left). The measurements began at 9:40 am on the 13$^{\rm th}$ September 2022.
({\bf C})
Variation over time of the bedform height, position and length (three lower panels), as measured along the central longitudinal transect (dashed line in {\em B}). Solid lines are obtained using the model with best fit parameters to the data: $\{L_0/L_{\rm sat},\, q_{\rm in}/q_{\rm sat}^{\rm e}\}=\{2.0,\,2.0\}$, see also Fig.~\ref{fig: fit}. During the measurement period, the wind was very consistent in strength (top panel) and direction ($75^{\rm o}$ anticlockwise from North, see SI Fig.
S14).
}
\label{fig: field}
\end{figure}

\section{Meter-scale sand bedforms in the Namib Desert}
\label{Sec:Field}
Field measurements were collected of sandy bedforms migrating over a gravel surface at Helga’s Interdune, Gobabeb, Namibia (Fig.~\ref{fig: field}A). This location has a wide interdune surrounded by linear dunes to the east and west, and a crossing dune to the south. Measurements of growing and migrating bedforms were undertaken during an easterly wind on the 13$^{\rm th}$ September 2022 \cite{Nield-data2023}, whilst measurements of shrinking bedforms were undertaken during an easterly wind on the 12$^{\rm th}$ September 2023 \cite{Nield-data2025a}. Grain size distributions from samples collected on these large dunes, as well as in traps located next to the small bedforms, support the hypothesis that the sand feeding these bedforms comes from the dunes (SI Fig.~S18). Surface topography was measured using a Leica P20 and P50 Scanstation for 2022 and 2023 datasets respectively, measuring at a horizontal resolution of $3.1$~mm at $10$~m. During both measurement periods, bedforms from within a larger field of similar forms (Fig.~\ref{fig: field}A) were chosen for further analysis because they were close to the Terrestrial Laser Scanner (TLS) and with its sight line perpendicular to the wind direction, in order to minimize occlusion of the underlying surface by the saltation cloud. Near-surface wind speed and direction within the bedform field were recorded on the consolidated bed at a height of $0.24$~m and  frequency of $10$~Hz using a Campbell CSAT 3D sonic anemometer located $\simeq 50$~m away from the bedform.

Each TLS scan was filtered for saltating grains using a radial filter (35$^{\circ}$ angle) and subsequently identified surface points were gridded at $0.01$~m resolution using the methods described in \cite{Niel11b}. Underlying topography was detrended for slope by fitting a surface through points where the surface change between all scans was less than $2.0 \times 10^{-4}$~m. This methodology gives a resolution of $5.5 \times 10^{-4}$~m for vertical change and a horizontal resolution of $0.01$~m \cite{Badd18}. The detrended surface was also smoothed using a $0.45 \times 0.45$~m$^2$ mean moving window filter to remove ripples \cite{Badd18}. These detrending and smoothing data processes are illustrated in SI Fig.~S17. Profiles $h(x)$ were then extracted from the centreline of individual bedform Digital Elevation Models (DEM) at successive times (Fig.~\ref{fig: field}B).

For each of these profiles, the height $H$ and position of the crest of the bedform can be deduced by fitting a parabola around the maximum of the elevation. Similarly, the upwind and downwind edges of the bedform can be determined, but this is a less precise measurement because of data noise where the elevation is small. A more robust method to estimate bedform length is to compute the area $\Delta = \int h(x) {\rm d}x$ between these two edges and define the length as $2\Delta/H$. The advantage here is that this triangle-like definition is less sensitive to the precise location of these edges, as the corresponding side regions do not contribute substantially to the integral. Note that, when dealing with theoretical profiles only, i.e. not related to data analysis or fitting, the ordinary definition of bedform length (difference between edge positions) is used instead.

For the bedform whose DEM evolution is displayed in Fig.~\ref{fig: field}B, we thus obtain its height, position and length as a function of time, showing a clear growth and migration, while its length remain fairly constant over the observation duration (Fig.~\ref{fig: field}C). These are the data against which the performance of our model (described below) is tested.

\begin{figure}[p]
\centering
\includegraphics[width=0.85\linewidth]{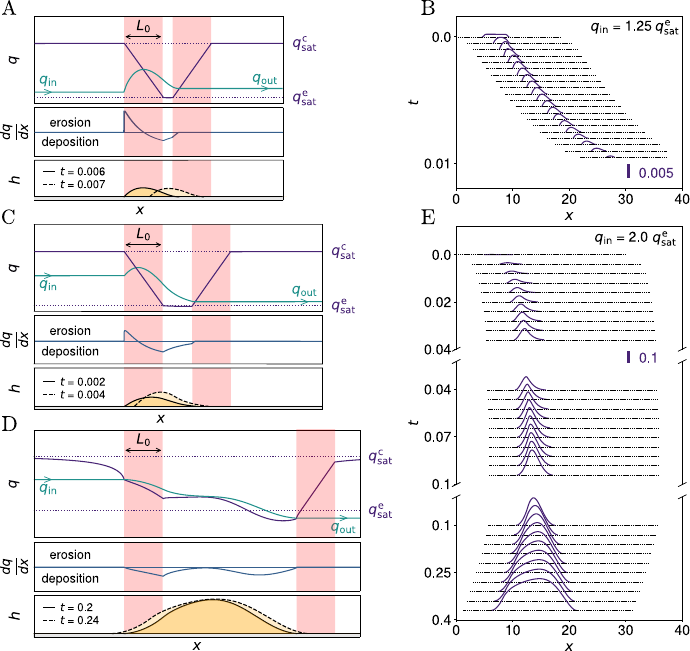}
\caption{
{\bf Three differing dynamical regimes of the bedforms as predicted by the model.}
In the left column, various profiles are schematized: elevation $h$ (black), saturated $q_{\rm sat}$ (violet) and actual $q$ (green) fluxes, and erosion rate $dq/dx$ (blue). Solid and dashed black lines correspond to $h$ at two successive times (legend), but the flux profiles are represented at the first time only. Transition zones of the transport capacity, of size $L_0$, are highlighted in pink. Right column: corresponding spatio-temporal diagrams of $h$ (reference amplitude bar in inset) are displayed for a given value of the input flux $q_{\rm in}$ (legend), see also SI Fig.
S2. All lengths in units of $L_{\rm sat}$ and times in units of $L_{\rm sat}^2/Q$. For all these simulations, the model parameters are $L_0=2$, $L_{\rm i}=5$, and $H_{\rm i}=10^{-3}$.
({\bf A, B})
Disappearing bedforms. For low values of $q_{\rm in}$, the sharp increase in the sand flux on the upwind part of the patch is not compensated for by the downwind sand flux relaxation which results in $q_{\rm out}>q_{\rm in}$.
({\bf C})
Growing and migrating bedforms. For larger $q_{\rm in}$, the sand flux has more space to relax to a value closer to $q_{\rm sat}^{\rm e}$, resulting in $q_{\rm out}<q_{\rm in}$.
({\bf D})
Spreading bedforms. The bedform eventually stops migrating and continues to grow both upwind and downwind. This last regime occurs when the shape is steep enough to significantly affect the wind flow and lower the shear stress (and thus $q_{\rm sat}$) on its upwind side. See also SI Fig.
S1.
({\bf E})
The spreading phase follows the migration phase as the bedform grows, and its stoss side becomes steeper. 
}
\label{fig: regimes}
\end{figure}

\section{A model for the dynamics of meter-scale sand bedforms}
\label{Sec:Model}
The model proposed herein for reproducing the dynamics of meter-scale sand bedforms is based on the core of well-established dune continuum approaches \cite{Kroy02a,Kroy02b,Andr02a,Hers04a,Hers04b,Dura10,Char13,Cour15}. First, we present the general form of the governing equations that relate the bed elevation profile $h$, saltation flux $q$ and bed shear stress $\tau$. We then introduce two additional key components to account for the difference in transport capacity between erodible and consolidated beds \cite{Neum98,Ho11,Delo23}. These are distinct maximum transport rates and, most importantly, a new length scale, $L_0$, which governs changes in transport capacity following shifts in surface properties on heterogeneous beds. For the sake of simplicity, the model is assumed to be homogeneous in the direction transverse to the wind, and we consider only the horizontal wind direction $x$ and the vertical axis of the bed elevation profile. This will be relevant for the dynamics over time $t$ of the central transect of the bedform, which we assume to be representative of the entire object.

\subsection*{Governing equations of dune theory}
\label{Sec:Model-Eqs}
The first equation is mass conservation, which stipulates that the erosion/deposition rate must balance the spatial variations in sediment flux:
\begin{equation}
    \partial_t h + \partial_x q=0.
    \label{MassConservation}
\end{equation}
$q$ is here defined as a volumetric flux: it counts the volume of the grains passing through a vertical surface of unit transverse width per unit time. This volume is taken at the bed packing fraction, thus eliminating any additional factors in Eq.~\ref{MassConservation}, also referred to as the Exner equation \cite{Exne20,Exne25}. As we consider a finite amount of sand deposited on a consolidated bed, the variations in flux must account for the fact that no erosion occurs below the reference level $h=0$ (see Methods).

The second component of the model differentiates between the flux $q$ and the transport capacity, or saturated sand flux, $q_{\rm sat}$. These two quantities coincide in the steady state case of a homogeneous flat bed, but are generally distinct. Here we assume that $q$ tends to $q_{\rm sat}$ following a first-order relaxation process:
\begin{equation}
   L_{\rm sat}\partial_x q = q_{\rm sat}-q.
   \label{RelaxationLsat}
\end{equation}
This equation involves a key spatial scale, the saturation length $L_{\rm sat}$, which encodes the typical lag by which $q$ is delayed with respect to $q_{\rm sat}$ \cite{Saue01,Andr02a,Elbe05,Andr10,Dura11,Char13,Cour15,Paht17,Lamm17,Selm18,Lu21}. Because of the non-erodible reference level $h=0$, this equation cannot be satisfied if it would lead to erosion of the consolidated bed. To illustrate this point, consider the simple situation where saltation occurs over a perfectly flat consolidated bed at a transport rate $q$ that is below its transport capacity $q_{\rm sat}$. In this case, Eq.~\ref{MassConservation} would predict erosion. However, to keep $q$ constant, we impose $\partial_x q = 0$, thereby ensuring that the bed elevation remains at $h=0$.

To account for changes in transport capacity, $q_{\rm sat}$ is an increasing function of the bed shear stress, $\tau$, which itself is modified by the topography. In the limit of flat bedforms, as those considered here, the relationship between the stress perturbation and $h$ is better expressed in the Fourier space where all functions are decomposed over wavelength $\lambda$ or wavenumber $k=2\pi/\lambda$ ($\hat{f}$ will denote the Fourier transform of $f$). It reads
\begin{equation}
    \hat{\tau} = \tau_0 (\mathcal{A}+i\mathcal{B}) k \hat{h},
    \label{AandB}
\end{equation}
where $\tau_0 = \rho_{\rm air} u_*^2$ is the reference stress over the flat bed, $\rho_{\rm air}$ is the air density, and $u_*$ is the wind shear velocity. The two dimensionless coefficients $\mathcal{A}$ and $\mathcal{B}$ represent respectively the in-phase and in-quadrature responses of the shear stress to the bed modulation. Their values generally depend on $k$, but they can be taken as constants for turbulent flows over rough surfaces, which is the relevant limit for terrestrial winds over sand beds on which saltation occurs \cite{Char13,Cour24}. Here, we will take $\mathcal{A}=3$ and $\mathcal{B}=1.5$ as representative values of the measurements taken in the field \cite{Clau13,Lu21}.

\subsection*{Transport capacity on erodible and consolidated beds}
\label{Sec:Model-evsc}
The transport law, i.e. the transport capacity in homogeneous and steady conditions, that relates the saturated flux to the wind shear stress depends strongly on the nature of the bed. In the reference case of saltation occurring on an erodible sandy surface, $q_{\rm sat}$ typically increases linearly with $\tau$ above a threshold $\tau_{\rm th}$ \cite{Unga87,Rasm96,Andr04,Dura11,Vala15}. Here we write it as
\begin{equation}
    q_{\rm sat}^{\rm e}(\tau)=Q(\frac{\tau}{\tau_{\rm th}}-1),
    \label{TransportLawErodible}
\end{equation}
where $Q$ is a dimensional constant (in m$^2$\,s$^{-1}$), which accounts for dependencies of the sand flux on environmental parameters (grain size $d$ and density $\rho_{\rm sed}$, air density $\rho_{\rm air}$, gravitational acceleration $g$). All the values of these parameters are set below in order to compare the predictions of the model to the field data. As for the general theoretical analysis of the model, however, the results are displayed with lengths in units of $L_{\rm sat}$ and time in units of $L_{\rm sat}^2/Q$.

The linear relationship between transport capacity $q_{\rm sat}$, and bed shear stress, $\tau$, observed in steady-state conditions can be attributed to the fact that the majority of the moving grains contributing to the flux remain close to the surface, below Bagnold's focal height. As a result of the feedback of these grains on the airflow, as well as the rebound condition on a sandy bed, their velocity is independent of the wind shear stress and equal to the transport threshold \cite{Andr04,Dura11,Crey09,Vala15}. For significantly large winds, typically above a Shields number $\Theta = \tau/(\rho_{\rm sed} g d) \simeq 0.3$, a quadratic correction to this linear behavior has been proposed, associated with binary particle collisions \cite{Paht20,Rala20}. Here, our field data are well below ($\Theta \simeq 0.05$) and we can thus safely neglect this correction. On the other hand, when saltation occurs over a consolidated surface, the rebounds of the grains on the bed are less dissipative and the transport layer is substantially thicker, with a vanishing feedback of moving grains on the flow. Their velocity then scales with that of the flow, i.e. $\sim \sqrt{\tau/\rho}$, which gives then an extra factor in the saturated flux \cite{Ho11}. Here, we thus write this second transport law as
\begin{equation}
    q_{\rm sat}^{\rm c}(\tau)=Q_{\rm c}\sqrt{\frac{\tau}{\tau_{\rm th}}}\left(\frac{\tau}{\tau_{\rm th}}-1\right),
    \label{TransportLawConsolidated}
\end{equation}
where we keep the same threshold shear stress $\tau_{\rm th}$ as in Eq.~\ref{TransportLawErodible}. This is a good approximation, and well supported by the few data available for comparison of the two transport laws \cite{Ho11,Kama22,Delo23}. In contrast, determination of the ratio $Q_{\rm c}/Q$ from these data is less straightforward as it is found to vary from $1$ to $3$ depending on the conditions and measurement set-up. For the sake of simplicity, we set here $Q_{\rm c}=Q$, but bear in mind that this ratio is, in principle, an additional adjustable parameter of the model.

In fact, the precise parametric choices for these two transport laws do not change the results qualitatively, as long as the transport capacity is enhanced over a consolidated bed, as compared to an erodible granular surface. Much more critical, however, is the necessity to account for the transition from one law to the other, in order to reproduce the observed growth and migration of these small bedforms. This process is associated with a length scale, here denoted by $L_0$, that controls the change in transport capacity after a shift in bed surface properties. Such a non-abrupt transition is consistent with the observations of spatial variations of saltation over sand patches \cite{Niel11a,Delo23}. In the present work, $L_0$ is an empirical parameter, which is adjusted based on the measurements made in the field. Its dependence on grain/flow properties is part of the discussion, but is clearly beyond the scope of this paper and would require further dedicated studies at the microscopic scale. In the present analysis, we simply assume that the transport capacity undergoes a linear downwind transition when the bed switches from state~1 to state~2 (consolidated to erodible or vice-versa) at location $x_s$:
\begin{equation}
    q_{\rm sat} =
    \begin{cases}
    q_{\rm sat_1} & \text{if $x \leq x_s$}, \\
    q_{\rm sat_1} + \dfrac{x-x_s}{L_0}(q_{\rm sat_2}-q_{\rm sat_1})& \text{if $x_s \leq x \leq x_s+L_0$},\\
    q_{\rm sat_2} & \text{if $x \geq x_s+L_0$}.
    \end{cases}
    \qquad
    \label{eq: trans_L0}
\end{equation}
These expressions assume that the bed is in state~2 for more than $L_0$, otherwise the transition is shortened and the switch back to state~1 starts before reaching $q_{\rm sat, 2}$ (SI Appendix). One could argue that windward (from consolidated to erodible beds) and leeward (from erodible to consolidated beds) transition lengths should be different, but we have not explored this more refined option that requires an additional parameter. Several other options than the piece-wise linear form (\ref{eq: trans_L0}) have also been tested, including a smoother transition shape with a hyperbolic tangent, and a centered scheme with $q_{\rm sat}$ starting/ending its transition at $x_s \mp L_0/2$. Furthermore, the precise point at which the bed transitions from consolidated to erodible or vice-versa is not readily discernible. Numerical simulations have demonstrated that a gradual shift from consolidated to erodible states occurs when the surface is covered by a few grains of erodible sand \cite{Kama22}. Here, we set the switching criterion at a finite bed elevation $h_{\rm s} = 2\times 10^{-4}\;L_{\rm sat}$, typically corresponding to 1 or 2 grain diameters. These arbitrary choices made for the shape of the transition in transport capacity and the threshold bed thickness do not qualitatively modify the results presented below, but slightly affect the position and shape of the upwind and downwind edges of the bedform.

\begin{figure}[t]
\centering
\includegraphics[width=\linewidth]{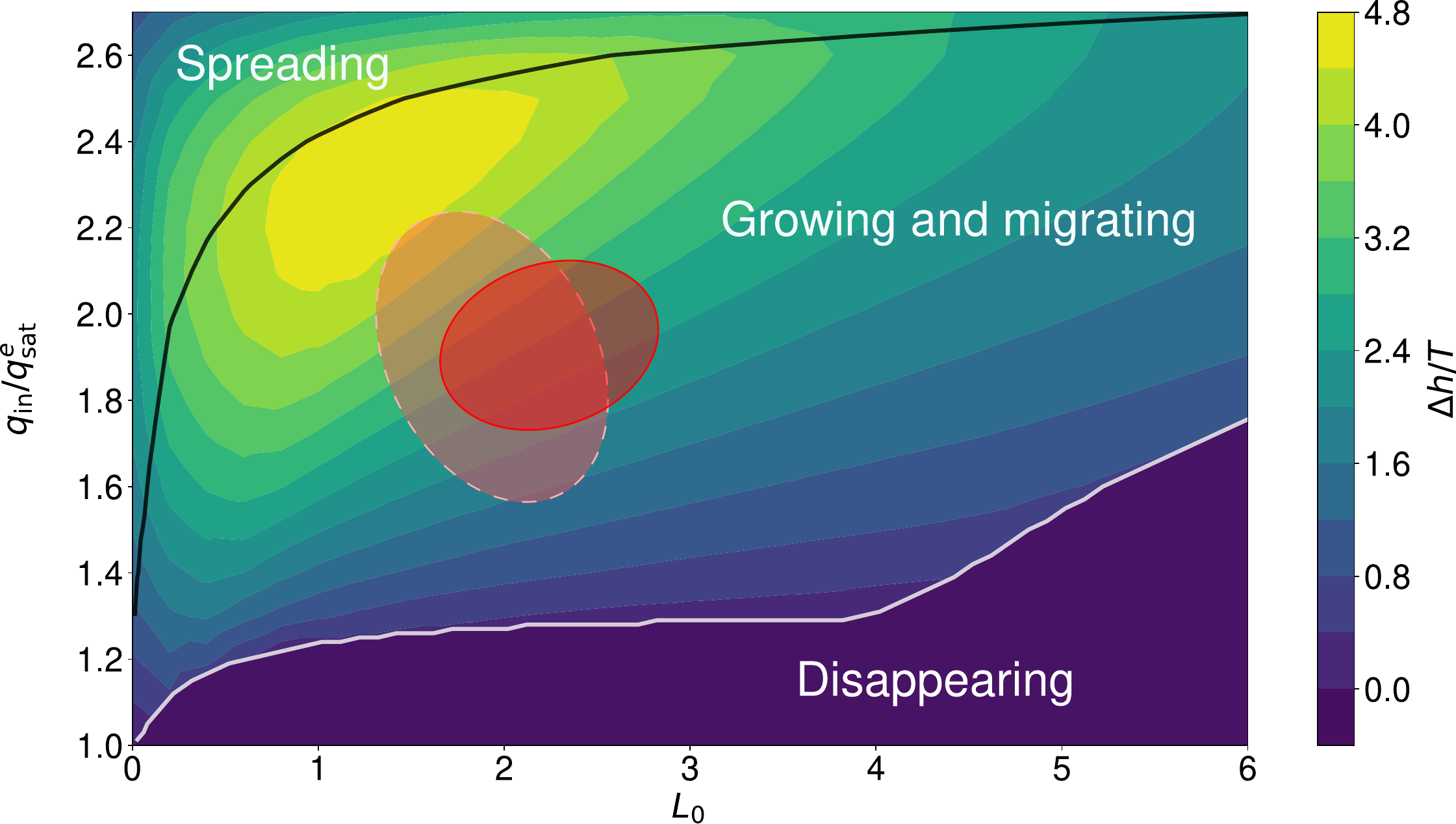}
\caption{
{\bf Diagrammatic representation of the three dynamical regimes in the parameter space $\{q_{\rm in}/q_{\rm sat}^{\rm e},\,L_0\}$ at a given time $T$.}
The color code indicates the average growth rate of a bedform at time $T = 0.02$, starting from a flat patch $t=0$ ($L_{\rm i}=5$, $H_{\rm i}=10^{-3}$). Black line: limit between migrating (below) and spreading (above) bedforms. White line: limit between growing (above) and disappearing (below) bedforms. The red (plain outline) and pink (dashed outline) areas show the estimated values (and their uncertainties) of input flux and transition length when the model is compared to the field data, see respectively fits in Fig.~\ref{fig: fit} (case corresponding to the red ellipse) and SI Fig.
S12 (pink ellipse). As can be seen in these figures, the time interval during which the growth and migration of the bedform was compared to the model was respectively $0.004$--$0.013$ (Fig.~\ref{fig: fit}) and $0.006$--$0.012$ (SI Fig.~S12) in dimensionless units, hence the choice of $T = 0.02$ for this diagram, as a relevant capping rounded number.
}
\label{fig: diagram}
\end{figure}

\section{Three different dynamics of bedforms evolution}
\label{Sec:ThreeDynamics}
We integrate the above equations numerically to simulate the time evolution of a bedform, starting from an almost flat patch of height $H_i$ and limited length $L_i$. We find that the bedform profile does not converge towards a propagative steady state, but displays three different dynamic regimes depending on the values of the input flux $q_{\rm in}$ and the transition length $L_0$: disappearance, growth and migration, and spreading. In practice, we have used the initial profile $h(x,t=0) = H_i \left( 1 - \cos (\pi x/L_i) \right)^\alpha$ for $0 \leq x \leq L_i$ and $h=0$ otherwise, where the exponent was chosen large enough ($\alpha = 40$) to make the shape of this profile very flat. We have tested other profiles, like a Fermi-Dirac shape, with little difference in the results. Unless otherwise stated, we have set the initial height to $H_i=5\,h_{\rm s}$ and length to $L_i=5\,L_{\rm sat}$.

\subsection*{Growth, migration and spreading}
In order to illustrate and elucidate these three dynamic regimes and their associated mass balances, Fig.~\ref{fig: regimes} displays the different longitudinal elevation profiles together with the flux and erosion rates. We first consider the most interesting case where the incoming flux lies between the transport capacities on consolidated and erodible beds: $q_{\rm sat}^{\rm c} > q_{\rm in} > q_{\rm sat}^{\rm e}$. The more ordinary case where $q_{\rm in}$ is smaller than $q_{\rm sat}^{\rm e}$ is discussed briefly at the end of this section.

The generic picture is as follows. When saltating grains move from a consolidated surface to an almost flat, sandy bedform, the transport capacity decreases from $q_{\rm sat}^{\rm c}$ to $q_{\rm sat}^{\rm e}$ over the length $L_0$ (Eq.~\ref{eq: trans_L0}), but the actual sand flux $q$ first increases. This is associated with erosion of the upwind edge of the bedform (Eq.~\ref{MassConservation}). The flux then decreases and sediment is deposited over the sand bed. It can also deposit beyond the downwind edge of the sandy area, where the ouput flux $q_{\rm out}$ is eventually released. Once again, this occurs because the switch towards the larger transport capacity on the consolidated bed is not immediate. Rather, the transport capacity increases gradually over the scale $L_0$. These processes of erosion and deposition on the windward and leeward margins induce a systematic migration of the bedform.

These dynamics can be investigated in more detail according to the value of $q_{\rm in}$. Slightly increasing the incoming flux above $q_{\rm sat}^{\rm e}$, we first have $q_{\rm out} > q_{\rm in}$, i.e. a negative mass balance for the bedform. Its upwind side becomes steeper, but it gradually shrinks and accelerates. It eventually disappears, retaining its asymmetrical shape (Fig.~\ref{fig: regimes} A and B). Increasing $q_{\rm in}$ further, the same asymmetry develops over a shorter time, but at some large enough value, the mass balance becomes positive: $q_{\rm out} < q_{\rm in}$. The bedform then grows and slows down as it continues to migrate (Fig.~\ref{fig: regimes} C and E for $t<0.04\;L_{\rm sat}^2/Q$). This growth occurs mainly in height, as the length of the bedform remains fairly constant. When this growth is sustained for a sufficient time, the bedform becomes steep enough to interact significantly with the wind flow (Eq.~\ref{AandB}). This reduces the bed shear stress around both the windward and leeward margins (SI Fig.
S1), so that sand deposition occurs upwind and downwind of the bedform (SI Fig.
S2). As a result, its length increases, its height saturates, and a spreading dynamics follows the growth and migration regime (Fig.~\ref{fig: regimes} D and E for $t>0.1\;L_{\rm sat}^2/Q$).

The whole sequence displayed in Fig.~\ref{fig: regimes}E runs faster for larger values of $q_{\rm in}$. Furthermore, because the erosion and deposition zones result from the competition between the space variations of $q_{\rm sat}$ due to $L_0$ and the relaxation of $q$ associated with $L_{\rm sat}$ (Eq.~\ref{RelaxationLsat}), the ratio of these two length scales also governs the occurrence of the different regimes (SI Figs.
S3,
S4,
S5).

\subsection*{Conditions for the three regimes}
Fig.~\ref{fig: diagram} summarizes the dominant dynamics in the different regions of the parameter space $\{q_{\rm in}/q_{\rm sat}^{\rm e},\,L_0\}$, after a given time $T=0.02\;L_{\rm sat}^2/Q$, always starting from the same initial condition. The disappearing regime is for low values of the input flux, with a systematic but rather moderate dependence on $L_0$. By contrast, the spreading regime concerns larger values of the input flux and its delimitation in this diagram shows a strong variation with respect to $q_{\rm in}$ at small $L_0$. In between, for moderate values of both $q_{\rm in}$ and $L_0$, growth and migration occur systematically. Not surprisingly, as discussed in the next section, this is where the majority of the field observations lie. 

Fig.~\ref{fig: diagram} is a generic diagram in the sense that it always shows the same organization of the different regimes, regardless of the other parameters of the model and the initial conditions. However, the exact location at which the transition occurs from one regime to another (white and black lines in Fig.~\ref{fig: diagram}) is somewhat sensitive to the initial length and height of the patch (SI Fig.
S6). It also depends on the total integration time $T$, especially when the bedforms evolves from the growing and migrating regime to the spreading regime (Fig.~\ref{fig: regimes}E). In this case, the transition curves between these two regimes corresponds to smaller values of $q_{\rm in}$ for larger values of $T$. Here, for Fig.~\ref{fig: diagram}, the choice of $T=0.02$ is justified by the fact that the fitted data, which we represent with the red and pink ellipses, correspond to a growth and migration of the bedforms between $0.004$ and $0.015$ in dimensionless units (see next section).

To clarify the role of the initial conditions in the development of the bedform, it is interesting to consider a perfectly flat ($H_i=0$) erodible surface of length $L_i$, for which an analytical solution of the model exists (SI Appendix). One can in particular compute, in the parametric plane $\{q_{\rm in},\, L_i\}$, the line separating the two regions where patches are gaining/losing mass for large/small values of $L_i$ derived from the condition $q_{\rm in}=q_{\rm out}$ (see Eqs.
11, 
17 and SI Fig.
S7). When starting from more realistic configurations of patches with a non-zero initial height, $H_i>0$, the evolution of the bedform presents different phases that are sensitive to the initial mass of sand. For example, when the initial length $L_i$ of the patch is small enough to be in the conditions for which its mass balance is negative, it may not eventually disappear as, while loosing mass, it first extends in length and then may reach the other region for which it can begin to gain mass and grow before its height has vanished. Conversely, an initially long patch first increases in height but does so while decreasing in length. If $q_{\rm in}$ is not large enough, the rate of shrinking can outweigh that of growth and the patch eventually disappears. These different examples are displayed in SI Fig.
S7. Importantly, as these analytics are derived for the perfectly flat case, it cannot be used to compare the model to the field data, but allows for a better understanding of the way in which the processes of erosion and deposition work in the model.

Finally we discuss the case for which the input flux is smaller than the transport capacity on the erodible bed: $q_{\rm in} < q_{\rm sat}^{\rm e}$. For this purpose, it is interesting to consider first steady propagative solutions of the model, computed when imposing $q_{\rm in} =  q_{\rm out}$ (both less than $q_{\rm sat}^{\rm e}$). Examples of such solutions are discussed in \cite{Andr02a} for a single transport law associated with an erodible bed, and are displayed in SI Fig.
S8 together with elevation profiles for transport laws that account for both erodible and consolidated beds. Interestingly, when transport capacity is sensitive to the bed nature, but with a vanishing transition length $L_0=0$, one obtains identical profiles to the case where the single transport law for an erodible bed is considered, with shapes very close to a symmetric cosine. For those cases, the bedform length is on the order of the cut-off scale of the dune instability $2\pi L_{\rm sat} \mathcal{A}/\mathcal{B}$ (typically $\simeq 10$~m), almost independently of $q_{\rm in}$. For $L_0>0$, the profiles take this characteristic asymmetric, more realistic, shape, and have a smaller length. However, in all cases, the heights are smaller for larger fluxes and vanish when $q_{\rm in} \to q_{\rm sat}^{\rm e}$. Importantly, these propagative solutions are unstable in the sense that an incoming flux even slightly different to the steady state value makes the bedform evolve away from the steady state solution: if $q_{\rm in}$ is a little too small/large, the bedform shrinks/grows. For initial flat patches, an input flux $q_{\rm in} < q_{\rm sat}^{\rm e}$ makes bedforms typically shrink and disappear, as illustrated in SI Fig.
S9. Conversely, to make the bedform survive and potentially develop into a dune when $q_{\rm in} < q_{\rm sat}^{\rm e}$, it must first have sufficiently accumulated sand during the growing phase $q_{\rm in} > q_{\rm sat}^{\rm e}$ to have reached the critical mass of the steady state solution.

\begin{figure}[p]
\centering
\includegraphics[width=0.7\linewidth]{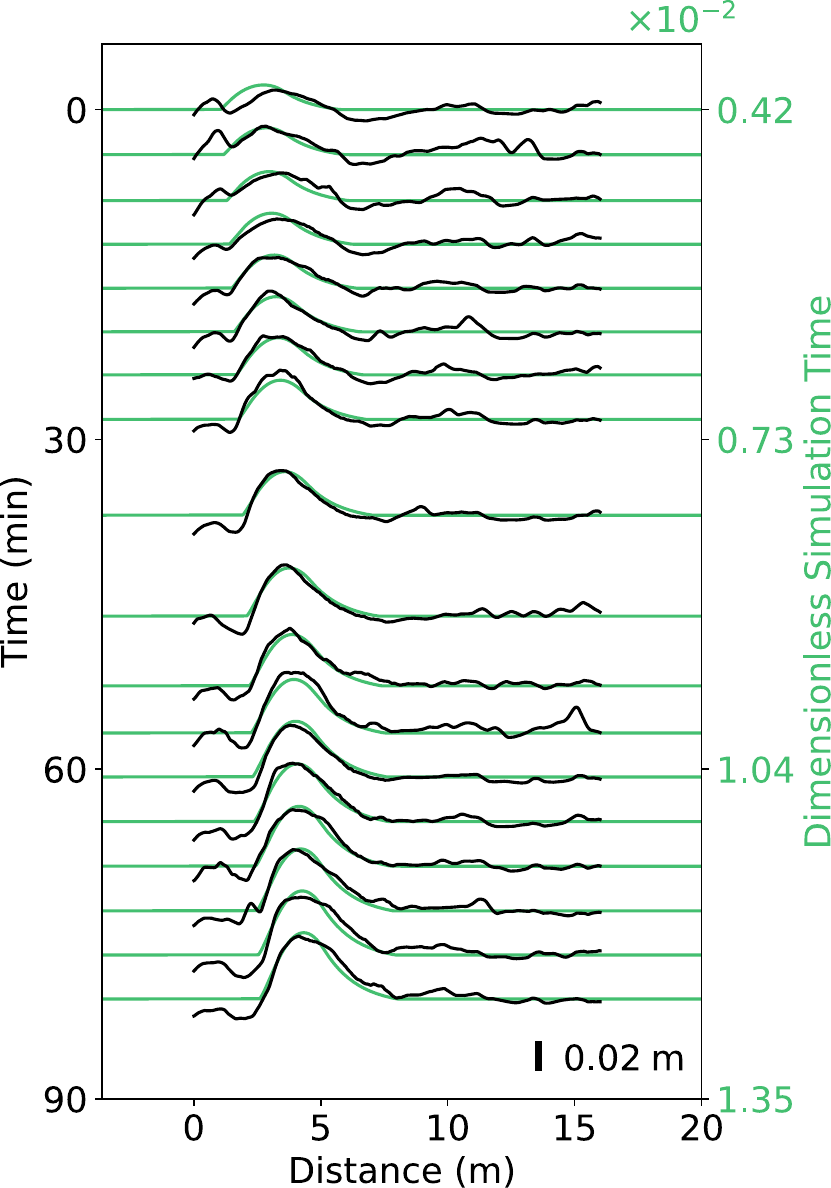}
\caption{
{\bf Model adjustment to field data.}
Black lines: spatio-temporal diagram with the elevation profiles of the central transect of the meter-scale bedform shown in Fig.~\ref{fig: field}B. Green lines: profiles from the model when the parameters are adjusted to the values $\{L_0/L_{\rm sat},\, q_{\rm in}/q_{\rm sat}^{\rm e}\}=\{2.0,\,2.0 \}$ that best fit the time evolution of bedform height, position and length (Fig.~\ref{fig: field}C). Time origin is 9:40 am. The ensemble of parameter values giving a reasonable fit of the data are displayed in Fig.~3 as a red ellipse. Right axis in green: dimensionless time in the simulation.
}
\label{fig: fit}
\end{figure}

\section{Comparison of the model to field data}
\label{Sec:ComparisonToData}
The model is able to reproduce the growth and migration of asymmetric sand bedforms similar to those observed in the field that occur when incoming flux is greater than the transport capacity over an erodible bed ($q_{\rm in}/q_{\rm sat}^{\rm e} \geq 1$). The high-resolution topographic data presented in Fig.~\ref{fig: field} provide a comprehensive spatio-temporal coverage of the evolution of such a bedform at the meter scale under a constant wind, which can be used to quantitatively test the theoretical predictions. As the model is two-dimensional, bedform height, length and position are computed over time from the elevation profiles along a central longitudinal transect (white dashed lines in Fig.~\ref{fig: field}B).

All the results of the theoretical analysis are given with lengths in units of $L_{\rm sat}$ and times in units of $L_{\rm sat}^2/Q$. For the comparison with the data (Fig.~\ref{fig: field}), we need to express these length and time scales for the specific values in the field where the measurements have been performed. We take a grain size $d=255\;\mu{\rm m}$, densities for the sediment $\rho_{\rm sed}=2.65\times 10^3\;{\rm kg\,m^{-3}}$ and the air $\rho_{\rm air}=1.2\;{\rm kg\,m^{-3}}$, the bed hydrodynamic roughness $z_0=10^{-3}\;{\rm m}$ and the gravitational acceleration $g=9.81\;{\rm m\,s^{-2}}$. We assume that the saturation length is well estimated by $L_{\rm sat}=2.2\,d\rho_{\rm sed}/\rho_{\rm air}=1.24\;{\rm m}$ \cite{Andr10}, and the threshold shear velocity by $u_{\rm th}=0.082\sqrt{gd\rho_{\rm sed}/\rho_{\rm air}} \simeq 0.19\;{\rm m\,s^{-1}}$ \cite{Cour24}. For the wind velocity measured during the bedform evolution (Fig.~\ref{fig: field}C), this corresponds to a velocity ratio $u_*/u_{\rm th}=2.8$, or $\tau/\tau_{\rm th}=7.84$ in terms of shear stress. Similarly, the sand flux constant $Q= 8.3 \frac{\rho_{\rm air} u_{\rm th}^3}{\rho_{\rm sed} g} \simeq 2.6\times10^{-6}\;{\rm m^2\,s^{-1}}$ \cite{Cour24}, which combined with the saturation length gives the characteristic time scale $L_{\rm sat}^2/Q=5.9\times10^5\;{\rm s}\simeq 7 \;{\rm days}$. Finally, with the two transport laws (\ref{TransportLawErodible},\ref{TransportLawConsolidated}), we obtain saturated sand fluxes over erodible $q^{\rm e}_{\rm sat} \simeq 1.8\times10^{-5}\;{\rm m^2\,s^{-1}}$ and consolidated $q^{\rm c}_{\rm sat} \simeq 5.0\times10^{-5}\;{\rm m^2\,s^{-1}}$ beds.  

The primary parameters of the model that require adjustment to fit the field data are the incoming sand flux $q_{\rm in}$ and the transition length $L_0$. Starting the simulation with a given almost flat patch, $q_{\rm in}$ and $L_0$ are tuned to reproduce the time evolution of the bedform height, length and displacement (green lines in Fig.~\ref{fig: field}C). This tuning results in full elevation profiles that are remarkably well-adjusted, with a quantitatively capture of the shape, growth and migration of the sand deposit over time (Fig.~\ref{fig: fit}). In fact, it is also necessary to adjust two other quantities, the time and space at which the comparison to the measured profiles begins. However, these are inconsequential offsets, essentially associated with the arbitrary choice of the initial patch height and length at time $t=0$. They do not affect the optimum $q_{\rm in}$ and $L_0$ to fit the data (SI Figs.~
S10 and 
S11). The values of these two parameters are also robust with respect to the criterion used to distinguish consolidated and erodible beds (the value of $h_{\rm s}$), as well as the choice for the transition profile of the transport capacity (Eq.~\ref{eq: trans_L0}). Furthermore, it was verified that the numerical integration steps do not influence the results as soon as they are sufficiently small (${\rm d}t \leq 10^{-6}$ with ${\rm d}x = 10^{-3}$).

Accounting for these various sources of uncertainty, the fitting process of the data set displayed in Fig.~\ref{fig: field} gives $q_{\rm in}/q_{\rm sat}^{\rm e} \simeq 2.0$ and $L_0/L_{\rm sat} \simeq 2.0$, with error bars on the order of $25$\%. These ranges of values are represented in Fig.~\ref{fig: diagram} by ellipses centered on the main values. As expected, they belong to the region of the parameter space associated with the growth and migration of bedforms. The same analysis performed on another bedform that emerged $30$ minutes later in the same area and under the same wind conditions gives consistent predictions (see second ellipse in Fig.~\ref{fig: diagram} and elevation profiles in SI Fig.
S12).

We have also run the model to reproduce the dynamics of a shrinking bedform similar to that of Fig.~\ref{fig: field}B, after the wind has started to change direction and reduce strength, yet still above transport threshold. The data show the evolution of elevation profiles that keep similar asymmetric shapes, but whose height and length decrease in time (SI Fig.
S13). We have accounted for the fact that the wind reduces with time, inducing a decreasing $q_{\rm sat}^{\rm e}$. The data can be fitted with a input flux proportional but significantly smaller than $q_{\rm sat}^{\rm e}$ and, perhaps more surprisingly, with a vanishing $L_0$. One can speculate on the relevance of any transition distance at low transport, but this case clearly requires more investigation.

Finally, SI Fig.~S16 provides field evidence of the third spreading regime, where we observed sand accumulation at the upwind side of the bedform over a period of approximately 2 hours \cite{Nield-data2025b}. However, these data are too sparse to be used for model parameter fitting.

\section{Discussion and perspectives}
\label{discussion}
The quantitative agreement between the theoretical predictions of the present model and the field measurements provides compelling evidence to explain the growth and migration of meter-scale bedforms, i.e. that are smaller than the minimum dune size predicted by linear stability analysis of a flat sand bed \cite{Andr02a,Char13,Lu21}. Our analysis shows how the dynamics of these bedforms is governed by bed heterogeneities, which modify the aeolian transport capacity when the nature of the bed, whether erodible or consolidated, changes. The typical asymmetrical shape of the observed elevation profiles is representative of an overall depositional process, resulting from a strong incoming flux that is over-saturated for saltation on a sandy bed \cite{Ho11,Delo23}. At the same time, erosion also occurs at the upwind edge of the bedforms, which induces their migration. This requires a local increase in sand flux, which is only possible if the transport capacity does not drop sharply, but over some specific distance, when the bed transitions from a consolidated to an erodible configuration.

Given the necessity of having a large incoming flux $q_{\rm in} > q_{\rm sat}^{\rm e}$ to observe the growth of these meter-scale sand bedforms, one must first ask where all this sand is coming from. On beaches or in the presence of wet sand beds, one can argue that surface grains can dry out or be mobilized by impacts to erode, and gradually increase transport \cite{Coop99,Stry24}. In arid desert dune fields however, where the consolidated bed in the interdune is mainly composed of coarse grains or gravels, the output flux from a dune can hardly exceed the value of $q_{\rm sat}^{\rm e}$. In this case, the saltation flux must gradually increase as it travels over interdune areas, mobilising the sand grains trapped between the larger particles by impacts or wind shear stress. An alternative explanation is to consider ephemeral sand deposits from saltation streamers, which could provide a significant amount of sand for transport when they disappear. The dynamics of saltation flux in the interdune areas or on the beaches could then be more dynamic than expected and a key process in the life of small-scale bedforms. This certainly merits further field studies measuring the input and output fluxes of these bedforms. In this perspective, the model proposed herein provides theoretical predictions of the magnitude of these fluxes as a function of the observed dynamics. In places where field measurements are not possible, this approach provides a means of using small-scale aeolian bedforms, evolving on short timescales, to provide valuable constraints on atmospheric flows and the associated sand fluxes.

In terms of the physics of sediment transport, an essential component of the present model is the transition length, $L_0$, which governs the change in transport capacity when bed properties change from a consolidated state to an erodible state, or vice versa. One would like to gain understanding of the physical processes at work in this transition. Valid questions might include whether these physical processes could be interpreted in terms of grain rebounds \cite{Ho11,Kama22}, or associated with the saturation length, or whether there is a dependence of $L_0$ on wind velocity. Of course, the saturated flux $q_{\rm sat}$ that quantifies this transport capacity cannot be measured directly and, whilst we are able to measure the actual flux, it is very challenging to obtain complete flux profiles $q(x)$ over various heterogeneous beds. At a microscopic level, recording the trajectories of saltating grains along the transition between the two types of bed are also desirable data that could be collected in the field as well as in wind tunnels. Theoretical analysis of such profiles and trajectories would be a useful way to improve current understanding of the transport processes associated with this transition, and how they relate to $L_0$.

The predicted bedforms do not reach a steady state, but instead exhibit three transient dynamics: increasing in height while migrating, disappearing, or growing in place while spreading. They must continue to be documented in the field in order to provide a more comprehensive picture of their out-of-equilibrium behavior. Further high-precision spatio-temporal data similar to those acquired in the Namib Desert should be obtained, particularly for bedforms in the disappearing and spreading regimes, as well as for various wind/grain/surface conditions. Another interest is to follow the dynamics of these bedforms over longer times, ideally over their entire life. The aim is then to identify the conditions under which they could become large enough to transform into dunes \cite{Kocu92}, for example on moist surfaces of beaches \cite{Hage18, Hage20, IJze24}. By combining numerical predictions and observations, the first step will be to estimate the strength and duration of the wind events that generate the appropriate incoming flux conditions to reach the spreading regime. As the model predicts bedform disappearance when $q_{\rm in}/q_{\rm sat}^{\rm e} \leq 1.2$ (Fig.~\ref{fig: diagram} for $L_0/L_{\rm sat}$ between $1$ and $3$), all these studies will also help to quantify the ephemeral existence of meter-scale sand deposits, and link their occurrence to specific atmospheric episodes. Also, in addition to their small size, it is worth considering whether a new branch should be added to the classification of aeolian dunes \cite{Cour24} for such transient or ephemeral bedforms.

At the larger length scale of a population of these bedforms (Fig.~\ref{fig: field}A, top), the study of their interactions, mediated by the sand flux, and possibly their resulting spatial organization, as illustrated in \cite{Hage18, Hage20, IJze24} on a beach environment, is of primary interest. A deeper understanding of their collective dynamics would allow them to be better identified, observed and tracked in the field and on aerial imagery. As they can disappear rapidly in weak or multidirectional wind conditions, their occurrence and resilience to different wind regimes would provide essential indirect information on sediment transport in areas of limited sand availability.

Finally, our work also suggests the need to investigate the complementary configuration, where a delimited area of consolidated surface is present in the middle of a mostly erodible bed. This has been observed, for example, on the dunes of the White Sands (New Mexico, USA), where when sufficiently humid, the gypsum grains stick to each other, providing such a consolidated surface over which mobile grains can be transported by the wind. How can such `holes' of mobile grains develop and possibly move? Precise spatio-temporal data are needed to test these new ideas.

\section*{Acknowlegments}
We thank B. Andreotti, S. Courrech du Pont, C. Gadal, N. Bristow and TOAD project partners for discussions. CN acknowledges financial support from from the French National Research Agency (ANR-23-CE56-0008/EOLE), the UnivEarthS LabEx program (ANR-10-LABX-0023), and the IdEx Universit\'e de Paris (ANR-18-IDEX-0001). Funding for the field component of this research is gratefully acknowledged from Natural Environment Research Council, UK and National Science Foundation, USA (NE/R010196/1 NSFGEO-NERC; NSF-GEO-1829541; NSF-GEO-1829513).  The authors thank the Gobabeb Namib Research Institute and G. Maggs-Kolling and E. Marais for field support and site identification and S. Nangolo and R. Huck for field assistance.  We acknowledge the Namibia Ministry of Environment, Forestry and Tourism, NCRST and Namib-Naukluft National Park (permit RPIV00052018). Field data processing used the Iridis Southampton Computing Facility. Thanks to P. Morgan and the Southampton Geography and Environmental Science Technician Team for undertaking the grain size analysis. We are grateful for GSD fieldwork assistance from A. Valdez and F. Bunch (permit GRSA-2018-SCI-004). For the purpose of open access, the author has applied a Creative Commons Attribution (CC BY) licence to any Author Accepted Manuscript
version arising. We thank the referees and the editor for constructive reviews.

\section*{Author Contributions}
JMN, MCB and GFSW carried out the acquisition of data in the field, and performed their treatment.
CR, PD, CN and PC developed and studied the model. CR, CN and PC adjusted it to the field data.
PC, CN, JMN, GFSW, MCB, KC and JB contributed the initial conceptualization and funding acquisition.
All authors discussed the results and contributed to the writing of the manuscript.

\section*{Appendix A: Material and methods}
We provide here details on how the equations of the model are numerically integrated, and in particular how bed elevation $h$ and sand flux $q$ are incremented over time, especially when erosion/deposition is limited by the consolidated bed.

Starting from given initial conditions, one first computes at each time step the along wind $x$-profiles of:\\
$\bullet$ The basal shear stress with $\tau(x,t)= \tau_0 [ 1 + \mathcal{A} \, {\rm IFFT}(k \hat h) + \mathcal{B} \, \partial_x h ]$, where $\tau_0$ is its reference value in the base (flat) state. Following notations of the main text, $\hat{h}(k)$ is the (fast) Fourier transform (FFT) of $h(x)$, and IFFT denotes the inverse transformation. $\mathcal{A}$ and $\mathcal{B}$ are constants here set to $3$ and $1.5$ respectively \cite{Lu21}.\\
$\bullet$ The saturated flux $q_{\rm sat}(x,t)$ from the transport laws with the above profile of $\tau$, accounting for the nature of the bed (erodible vs consolidated, see Eqs.~\ref{TransportLawErodible},~\ref{TransportLawConsolidated}) as well as for the transition region at the switch from two bed states (Eq.~\ref{eq: trans_L0}). Importantly, the saturated flux must not be negative, and if this calculation gives $q_{\rm sat}<0$ in some space intervals, we instead set it to $0$ there. In practice, this only happens at very late states of the bedform evolution, when the bed elevation profile becomes very steep and for which other aspects of the model should be modified (e.g., introduction of a recirculation bubble \cite{Andr02a}). None of the present results have reached this point.\\
The flux is then deduced from the relaxation equation (Eq.~\ref{RelaxationLsat}), here integrated through an explicit first order scheme in space:
\begin{equation}
q(x,t)= \dfrac{q(x-{\rm d}x,t) + \tfrac{{\rm d}x}{L_{\rm sat}} \, q_{\rm sat}(x,t)}{1 + \tfrac{{\rm d}x}{L_{\rm sat}}}
\label{UpDateq}
\end{equation}
starting from $q=q_{\rm in}$ at the up-wind side of the integration domain. Simultaneously, the bed elevation profile is up-dated with an explicit first order scheme in time:
\begin{equation}
h(x,t+{\rm d}t)=h(x,t) - \frac{{\rm d}t}{L_{\rm sat}} \, \left( q_{\rm sat}(x,t)-q(x,t) \right)
\label{UpDateh}
\end{equation}
At each grid point, we check that one does not erode more sand than available, and if the above calculation gives $h(x,t+{\rm d}t)<0$, we set $h(x,t+{\rm d}t)=0$ and correct the sand flux as $q(x,t) = q(x-{\rm d}x,t) + \frac{{\rm d}x}{{\rm d}t} \, h(x,t)$, so that mass conservation (\ref{MassConservation}) is still satisfied -- but (\ref{UpDateq}) is not. This applies in particular to the up-wind part of the domain, before the bedform, where the bed is flat with no sand available for erosion ($h=0$), and where $q = q_{\rm in} < q_{\rm sat}^{c}$ must be kept constant. The code integrating these equations is available in \cite{Rambert-Code2025}.

Field datasets have been deposited in the NERC EDS National Geoscience Data Centre \cite{Nield-data2023, Nield-data2025a, Nield-data2025b}.

\newpage
\section*{Appendix B: Analytical expression for the flux profile over a flat sand patch}
\label{FluxFlatPatch}
In the case of a perfectly flat patch subjected to a constant wind shear stress, the model can be solved analytically in order to obtain the longitudinal flux profile. $L_{\rm i}$ is the length of the erodible region, which starts at $x=0$. Importantly, this idealized configuration is used for a better understanding of erosion and deposition processes associated with flux variations, not for the fit of bedform elevation profiles from the field data.

\subsection*{Long initial patch length}
\label{large_Li}
According to the linear form of the transition in transport capacity proposed in the main manuscript, and assuming $L_{\rm i} \geq L_0$, the longitudinal profile of the saturated flux can be expressed in a piece-wise manner in five regions:
    \begin{subnumcases}{q_{\rm sat} =}
     q_{\rm sat}^{\rm c}
     & \text{Region I, $x \leq 0$,} 
     \label{qsatregionI}
     \\
     q_{\rm sat}^{\rm c} - \Delta q_{\rm sat} \dfrac{x}{L_0}
     & \text{Region II, $0 \leq x \leq L_0$,}
     \label{qsatregionII}
     \\
     q_{\rm sat}^{\rm e}
     & \text{Region III, $L_0 \leq x \leq L_{\rm i}$,} 
     \label{qsatregionIII}
     \\
     q_{\rm sat}^{\rm e} + \Delta q_{\rm sat} \dfrac{x-L_{\rm i}}{L_0}
     & \text{Region IV, $L_{\rm i} \leq x \leq L_{\rm i}+L_0$,} 
     \label{qsatregionIV}
     \\
     q_{\rm sat}^{\rm c}
     & \text{Region V, $x\geq L_{\rm i}+L_0$,}
     \label{qsatregionV}
    \end{subnumcases}
where we have defined:
\begin{equation}
\Delta q_{\rm sat} = q_{\rm sat}^{\rm c} - q_{\rm sat}^{\rm e}.
\label{eq: dqsat}
\end{equation}
With these expressions for $q_{\rm sat}$, we can integrate Eq.~2 of the main manuscript in these five regions, with the additional condition that $q$ stays constant if erosion cannot take place on the consolidated bed. In region~I, the flux is simply equal to the incoming flux: $q(x) = q_{\rm in}$. In region~II, we obtain
\begin{equation}
q(x) = q_{\rm in} \, e^{-x/L_{\rm sat}} - \Delta q_{\rm sat} \dfrac{x}{L_0}
+ \left( q_{\rm sat}^{\rm c} + \Delta q_{\rm sat} \dfrac{L_{\rm sat}}{L_0} \right) \left(  1 - e^{-x/L_{\rm sat}} \right).
\label{qofxregionII}
\end{equation}
At the upwind end, $q(0) = q_{\rm in}$. At the downwind end of region~II, the flux is
\begin{equation}
q(L_0) = q_{\rm sat}^{\rm e} + \Delta q_{\rm sat} \frac{L_{\rm sat}}{L_0}
\left[ 1 - \left( 1 + \frac{L_0}{L_{\rm sat}} - \frac{q_{\rm in} - q_{\rm sat}^{\rm e}}{\Delta q_{\rm sat}} \frac{L_0}{L_{\rm sat}}  \right) e^{-L_0/L_{\rm sat}} \right] \qquad
\label{qofL0regionII}
\end{equation}
From these expressions, we can also deduce the upstream erosion zone of the patch, for which $q \leq q_{\rm sat}$ in this region II. Its length $L_{\rm e}$ is computed from the condition $q(L_{\rm e}) = q_{\rm sat}(L_{\rm e})$. Using the above expressions it solves into
\begin{equation}
\frac{L_{\rm e}}{L_{\rm sat}} = \ln \left( 1 + \frac{q_{\rm sat}^{\rm c} - q_{\rm in}}{\Delta q_{\rm sat}} \frac{L_0}{L_{\rm sat}} \right).
\label{Lerosion}
\end{equation}
Similarly, in region~III, we have
\begin{equation}
q(x) = q(L_0) \, e^{-(x-L_0)/L_{\rm sat}} + q_{\rm sat}^{\rm e} \left(  1 - e^{-(x-L_0)/L_{\rm sat}} \right).
\label{qofxregionIII}
\end{equation}
At $x=L_{\rm i}$, using Eq.~\ref{qofL0regionII}, we obtain
\begin{equation}
q(L_{\rm i}) = q_{\rm sat}^{\rm e} + \Delta q_{\rm sat} \frac{L_{\rm sat}}{L_0} \, e^{-(L_{\rm i}-L_0)/L_{\rm sat}}
\left[ 1 - \left( 1 + \frac{L_0}{L_{\rm sat}} - \frac{q_{\rm in} - q_{\rm sat}^{\rm e}}{\Delta q_{\rm sat}} \frac{L_0}{L_{\rm sat}}  \right) e^{-L_0/L_{\rm sat}} \right] \qquad
\label{qofLiregionIII}
\end{equation}
Finally, in region~IV, we have
\begin{equation}
q(x) = q(L_{\rm i}) \, e^{-(x-L_{\rm i})/L_{\rm sat}} + \Delta q_{\rm sat} \frac{x-L_{\rm i}}{L_0}
+ \left( q_{\rm sat}^{\rm e} - \Delta q_{\rm sat} \frac{L_{\rm sat}}{L_0} \right) \left(  1 - e^{-(x-L_{\rm i})/L_{\rm sat}} \right).
\label{qofxregionIV}
\end{equation}
Importantly, this expression is only valid up to $x=L_{\rm d}$, defined as the location where $q$ and $q_{\rm sat}$ are equal in region~IV. A similar calculation to that of $L_{\rm e}$ gives
\begin{equation}
\frac{L_{\rm d}}{L_{\rm sat}} = \frac{L_{\rm i}}{L_{\rm sat}} + \ln \left( 1 + \frac{q(L_{\rm i}) - q_{\rm sat}^{\rm e}}{\Delta q_{\rm sat}} \frac{L_0}{L_{\rm sat}} \right).
\label{Ldeposition}
\end{equation}
For the rest of region~IV, and also in region~V, the flux stays constant as no erosion of the consolidated is possible. This allows to define the output flux as
\begin{eqnarray}
q_{\rm out} & = & q(L_{\rm d}) = q_{\rm sat}(L_{\rm d}) 
\nonumber \\
& = & q_{\rm sat}^{\rm e} + \Delta q_{\rm sat} \frac{L_{\rm sat}}{L_0} \, \ln \left( 1 + \frac{q(L_{\rm i}) - q_{\rm sat}^{\rm e}}{\Delta q_{\rm sat}} \frac{L_0}{L_{\rm sat}} \right),
\qquad
\label{qout}
\end{eqnarray}
where we have substituted the expression (\ref{Ldeposition}) for $L_{\rm d}$ in that for the saturated flux (\ref{qsatregionIV}). For the estimate the mass balance of such a flat patch, we must compare $q_{\rm out}$ to $q_{\rm in}$. With the above expression and inserting the expression of $q(L_{\rm i})$ (\ref{qofLiregionIII}) in the above of $q_{\rm out}$, the condition $q_{\rm out} = q_{\rm in}$ gives the following limiting curve in the parametric plane $q_{\rm in}$ \emph{vs} $L_{\rm i}$:
\begin{equation}
\frac{L_{\rm i}}{L_{\rm sat}} = \ln \frac{e^{L_0/L_{\rm sat}} - 1 - \frac{L_0}{L_{\rm sat}} + \frac{q_{\rm in} - q_{\rm sat}^{\rm e}}{\Delta q_{\rm sat}} \frac{L_0}{L_{\rm sat}} }{e^{\frac{q_{\rm in} - q_{\rm sat}^{\rm e}}{\Delta q_{\rm sat}} \frac{L_0}{L_{\rm sat}}} - 1} \, .
\label{qinvsLimassbalance}
\end{equation}
%

\subsection*{Short initial patch length}
\label{small_Li}
We now consider the case $L_{\rm i} \leq L_0$. The saturated flux profile then displays four regions only -- the central one disappears -- and we take:
    \begin{subnumcases}{q_{\rm sat} =}
     q_{\rm sat}^{\rm c}
     & \text{Region I, $x \leq 0$,} 
     \label{qsatregionIbis}
     \\
     q_{\rm sat}^{\rm c} - \Delta q_{\rm sat} \dfrac{x}{L_0}
     & \text{Region II, $0 \leq x \leq L_{\rm i}$,} 
     \label{qsatregionIIbis}
     \\
     q_{\rm sat}^{\rm c} + \Delta q_{\rm sat} \frac{x-2L_{\rm i}}{L_0}
     & \text{Region III, $L_{\rm i} \leq x \leq 2 L_{\rm i} $,} 
     \label{qsatregionIIIbis}
     \\
     q_{\rm sat}^{\rm c}
     & \text{Region IV, $x\geq 2 L_{\rm i}$.} 
     \label{qsatregionIVbis}
    \end{subnumcases}
In this case, the flux profile is the same in regions I and II as before, but its value at $x=L_{\rm i}$ is now
\begin{equation}
q(L_{\rm i}) = q_{\rm sat}^{\rm e} + \Delta q_{\rm sat} \frac{L_{\rm sat}}{L_0} \left[ 1 + \frac{L_0-L_{\rm i}}{L_{\rm sat}} \right.
- \left. \left( 1 + \frac{L_0}{L_{\rm sat}} - \frac{q_{\rm in} - q_{\rm sat}^{\rm e}}{\Delta q_{\rm sat}} \frac{L_0}{L_{\rm sat}}  \right) e^{-L_{\rm i}/L_{\rm sat}} \right]. \qquad
\label{qofLiregionIIIbis}
\end{equation}
In new region III, the flux is
\begin{equation}
q(x) = q(L_{\rm i}) \, e^{-(x-L_{\rm i})/L_{\rm sat}} + \Delta q_{\rm sat} \frac{x-L_{\rm i}}{L_0}
+ \left( q_{\rm sat}^{\rm c} - \Delta q_{\rm sat} \frac{L_{\rm i}+L_{\rm sat}}{L_0} \right) \left(  1 - e^{-(x-L_{\rm i})/L_{\rm sat}} \right). \qquad
\label{qofxregionIIIbis}
\end{equation}
If $L_{\rm i} \leq L_{\rm e}$ (given by Eq.~\ref{Lerosion}), then the whole patch is eroding. In this case, the output flux is $q_{\rm out} = q(L_{\rm i})$ (Eq.~\ref{qofLiregionIIIbis}), and $q_{\rm out} \geq q_{\rm in}$.

Conversely, when $L_{\rm i} \geq L_{\rm e}$, the profile (\ref{qofxregionIIIbis}) is valid until $q$ and $q_{\rm sat}$ are equal. The corresponding position $L_{\rm d}$ is
\begin{equation}
\frac{L_{\rm d}}{L_{\rm sat}} = \frac{L_{\rm i}}{L_{\rm sat}}
+ \ln \left[ 2 - \left( 1 + \frac{L_0}{L_{\rm sat}} - \frac{q_{\rm in} - q_{\rm sat}^{\rm e}}{\Delta q_{\rm sat}} \frac{L_0}{L_{\rm sat}} \right) e^{-L_{\rm i}/L_{\rm sat}} \right].
\label{Ldepositionbis}
\end{equation}
and the corresponding output flux is
\begin{equation}
q_{\rm out} = q_{\rm sat}^{\rm e} + \Delta q_{\rm sat} \frac{L_0-L_{\rm i}}{L_0} + \Delta q_{\rm sat} \frac{L_{\rm sat}}{L_0}
\ln \left[ 2 - \left( 1 + \frac{L_0}{L_{\rm sat}} - \frac{q_{\rm in} - q_{\rm sat}^{\rm e}}{\Delta q_{\rm sat}} \frac{L_0}{L_{\rm sat}} \right) e^{-L_{\rm i}/L_{\rm sat}} \right].
\label{qoutbis}
\end{equation}
The condition $q_{\rm out} = q_{\rm in}$ now gives an implicit relation between $q_{\rm in}$ and $L_{\rm i}$:
\begin{equation}
\frac{q_{\rm in} - q_{\rm sat}^{\rm e}}{\Delta q_{\rm sat}} \frac{L_0}{L_{\rm sat}} = \frac{L_0-L_{\rm i}}{L_{\rm sat}}
+ \ln \Bigg[ \, 2 - \left( 1 + \frac{L_0}{L_{\rm sat}} - \frac{q_{\rm in} - q_{\rm sat}^{\rm e}}{\Delta q_{\rm sat}} \frac{L_0}{L_{\rm sat}} \right) e^{-L_{\rm i}/L_{\rm sat}} \Bigg].
\label{qinvsLimassbalancebis}
\end{equation}
All the flux profiles corresponding to the above different cases are displayed in Fig.~\ref{Supplfig:FluxProfilesAnalytics}

\section*{Appendix C: Supplementary figures}
\label{SupplFigs}
\renewcommand\thefigure{S\arabic{figure}}
\setcounter{figure}{0}    

\begin{figure*}
\centering
\includegraphics[width=0.75\linewidth]{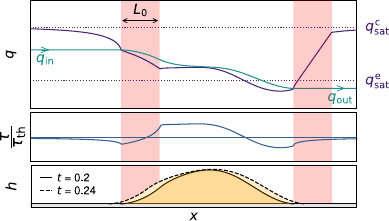}
\caption{
{\bf Shear stress profile for spreading bedforms.} Same as Fig.
2D of main manuscript, with the additional profile of the basal shear stress. Note the reduced $\tau$ both upwind and downwind of the bedform, due to the feedback of the topography on the flow (Eq.~3 of main manuscript). This also explains why $q(x)<q_{\rm sat}^{\rm e}$ downwind of the bedform.
}
\label{Supplfig:tauspreading}
\end{figure*}

\begin{figure*}
\centering
\includegraphics[width=0.75\linewidth]{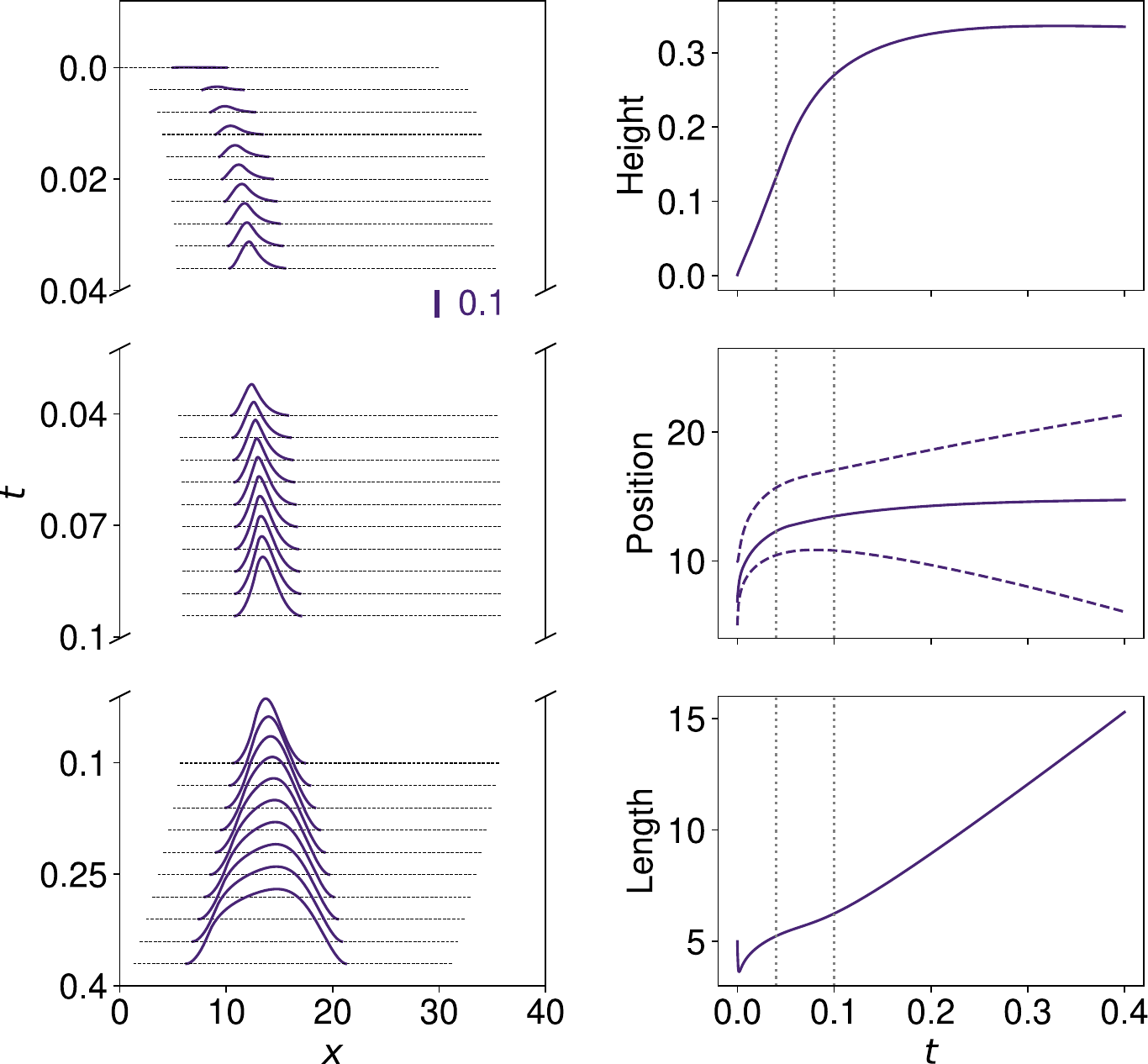}
\caption{
{\bf Time evolution of a growing bedform.} These graphs complement those of Fig.
2 of main manuscript, whose spatio-temporal diagram (panel E of that figure) is here reproduced on the left. On the right, we show the bedform height (top), the position of its crest (middle) as well as its upwind and downwind edges (represented by the dashed lines), and its length (bottom) as a function of time. Run for $\{L_0,\, q_{\rm in}/q_{\rm sat}^{\rm e}\}=\{2.0,\,2.0 \}$. The vertical dotted lines indicate the starting time of each sub-panel in the spatio-temporal diagram. Note that for $t>0.1$, the spreading regime is reached : the height remains constant while the upwind and downwind edges of the bedform move apart, thus increasing the length.
}
\label{Supplfig:patchlife}
\end{figure*}

\begin{figure*}
\centering
\includegraphics[width=0.75\linewidth]{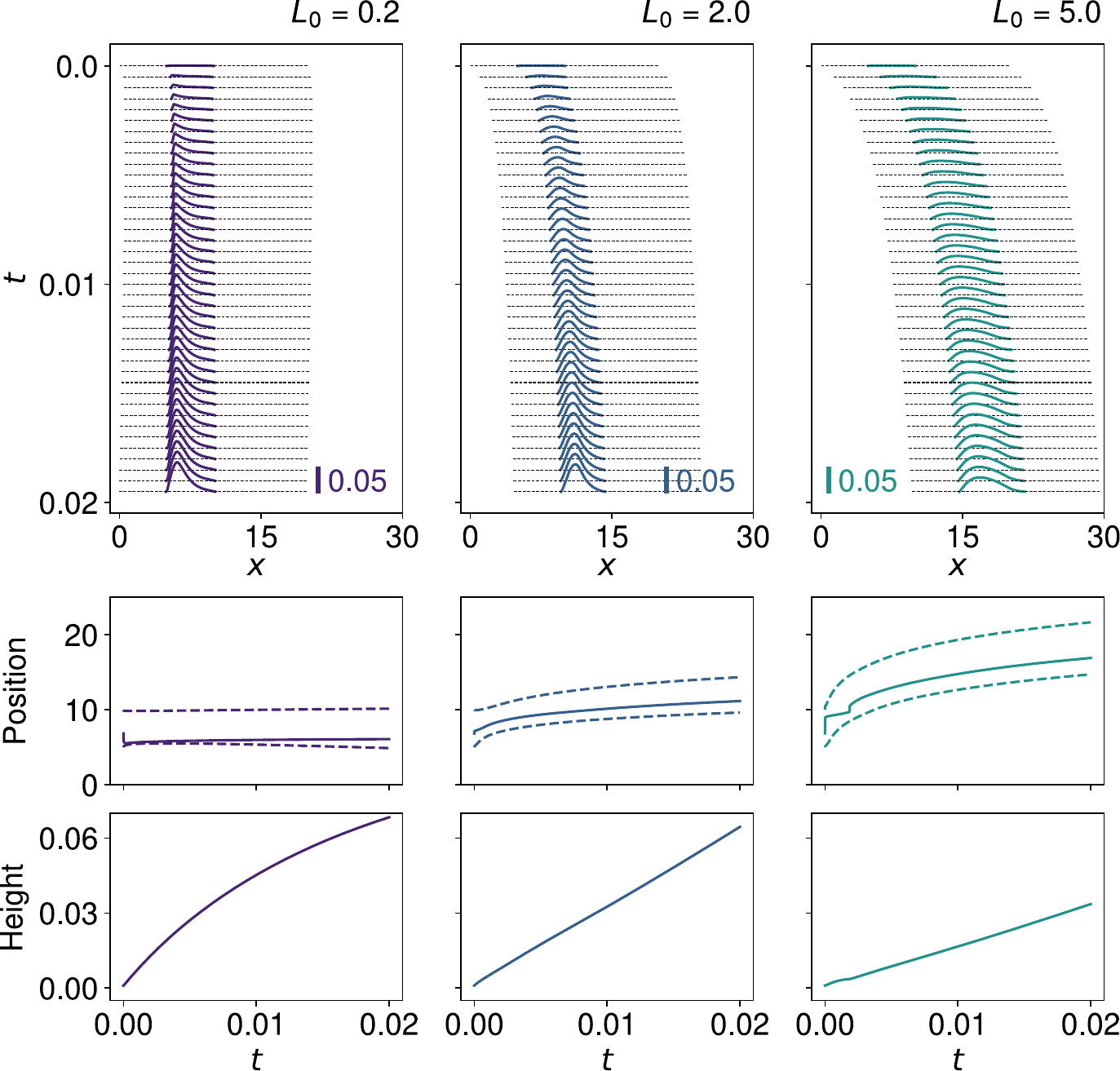}
\caption{
{\bf Influence of parameter $L_0$ on the evolution of the bedform.}
Top : Spatio-temporal diagrams. Middle : time evolution of the bedform crest (plain line) and edge positions (dashed lines). Bottom : time evolution of the bedform height.  Three values of $L_0$ are represented : 0.2 (left), 2 (middle) and 5 (right). All these runs are for $q_{\rm in} = 2\,q_{\rm sat}^{\rm e}$. For a given $q_{\rm in}$, the increase in $L_0$ leads to greater erosion at the upwind edge (see also Eq.~\ref{Lerosion} of SI text) and therefore reinforces the migration rate of the bedform. Furthermore, as $L_0$ increases, the bedform becomes less abrupt and its height decreases, as the sand is more evenly distributed over a greater distance.
}
\label{Supplfig:varyL0}
\end{figure*}

\begin{figure*}
\centering
\includegraphics[width=0.75\linewidth]{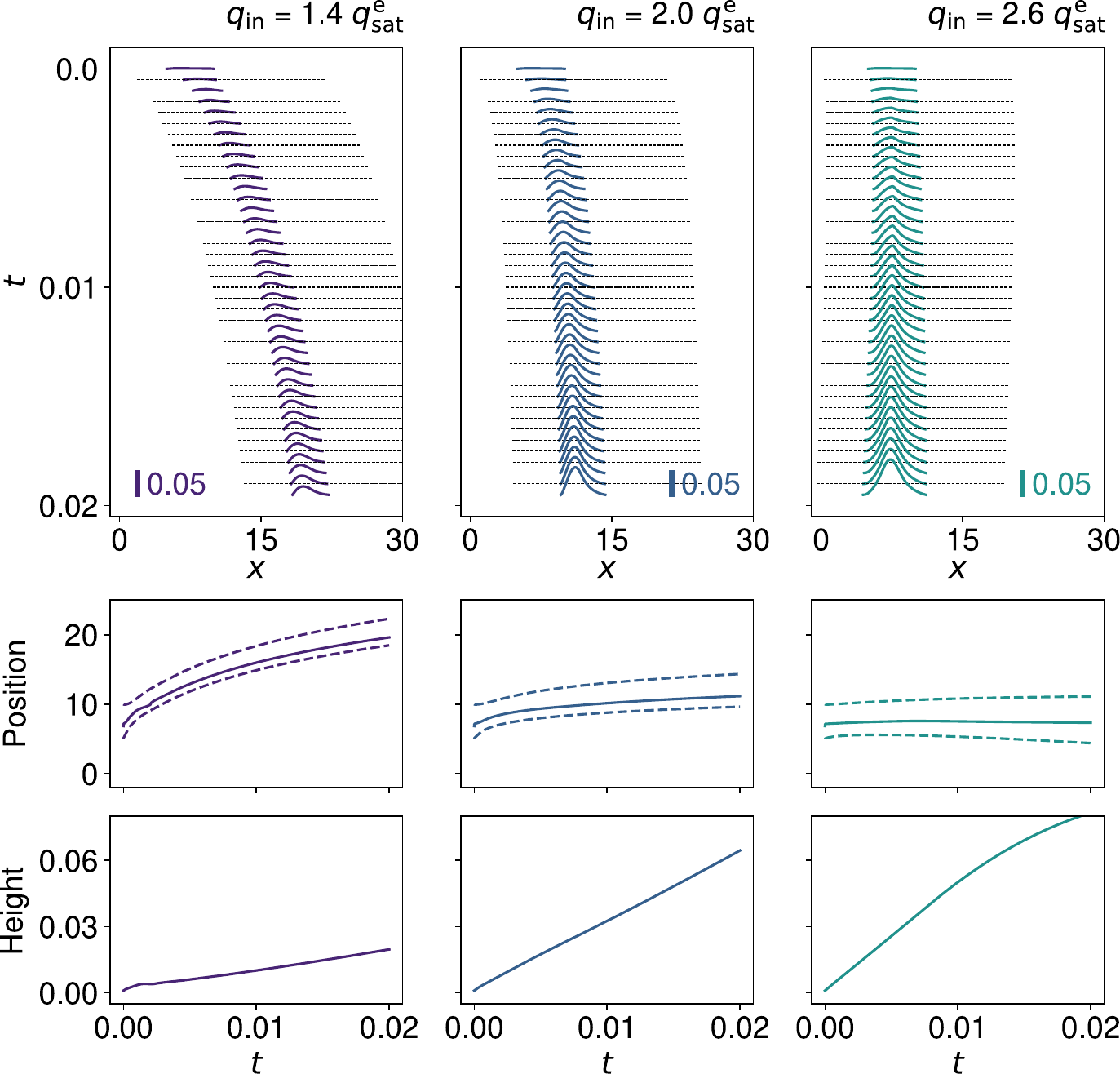}
\caption{
{\bf Influence of parameter $q_{\rm in}$ on the evolution of the bedform.}
Top : Spatio-temporal diagrams. Middle : Time evolution of the bedform crest (plain line) and edge position (dashed lines). Bottom : time evolution of the bedform height. Three values of $q_{\rm in}$ are represented : $1.4\,q_{\rm sat}^{\rm e}$ (left), $2\,q_{\rm sat}^{\rm e}$ (middle) and $2.6\,q_{\rm sat}^{\rm e}$ (right). All these runs are for $L_0 = 2$. For a given $L_0$, the increase in $q_{\rm in}$ leads to less erosion at the upwind edge (see also Eq.~\ref{Lerosion} of SI text) and therefore reduces the migration rate of the bedform. In addition, as $q_{\rm in}$ increases, the bedform height becomes greater, as more sand is deposited.
}
\label{Supplfig:varyqin}
\end{figure*}

\begin{figure*}
\centering
\includegraphics[width=0.75\linewidth]{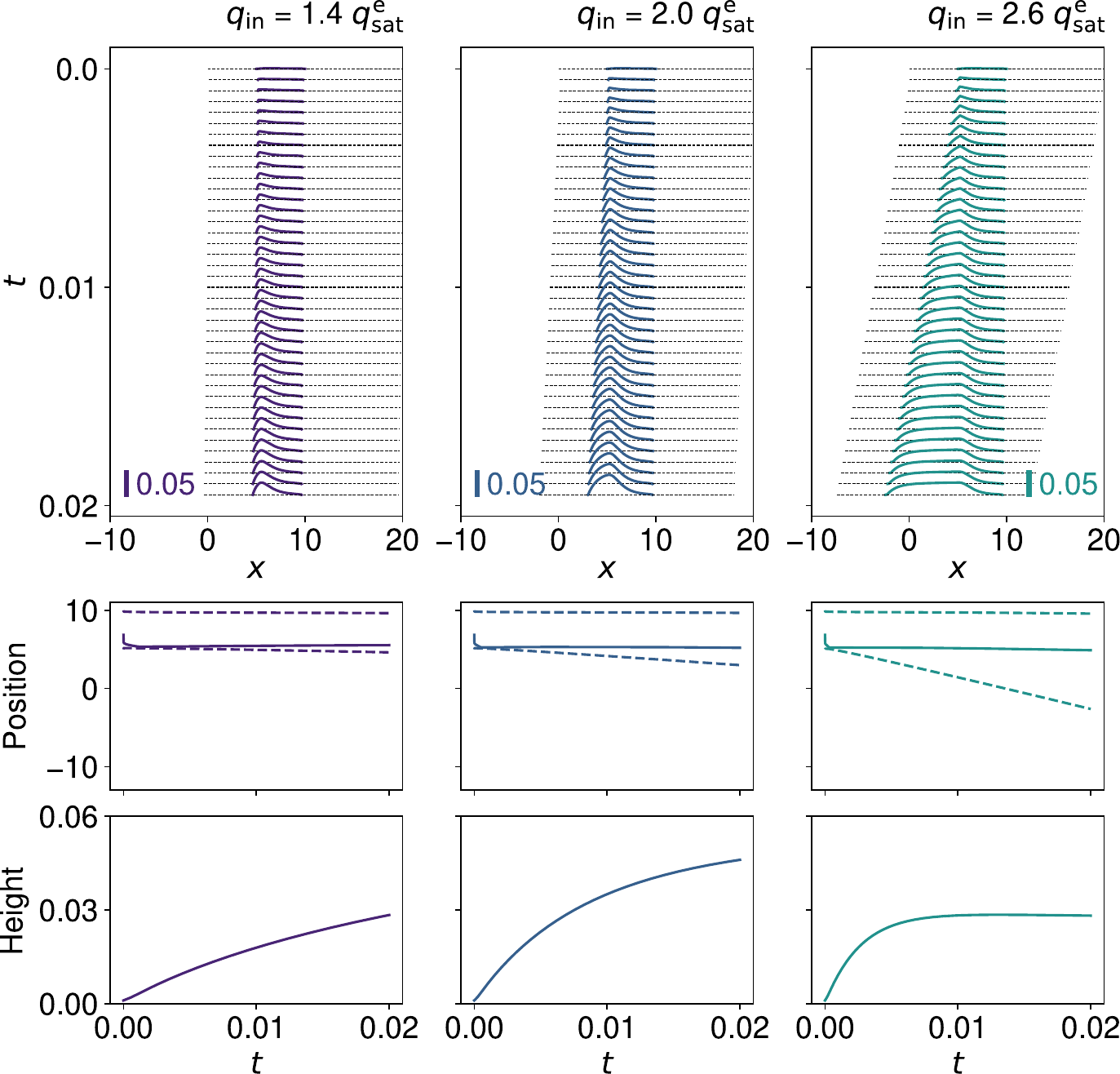}
\caption{
{\bf Spreading dynamics of the bedform for $L_0 = 0$.}
Same as Fig.~\ref{Supplfig:varyqin}, but for the special case $L_0=0$.
In this case, there is only deposition at the upwind edge. Increasing $q_{\rm in}$ leads to more deposition and therefore greater spreading.
}
\label{Supplfig:L0zero}
\end{figure*}

\begin{figure*}
\centering
\includegraphics[width=0.85\linewidth]{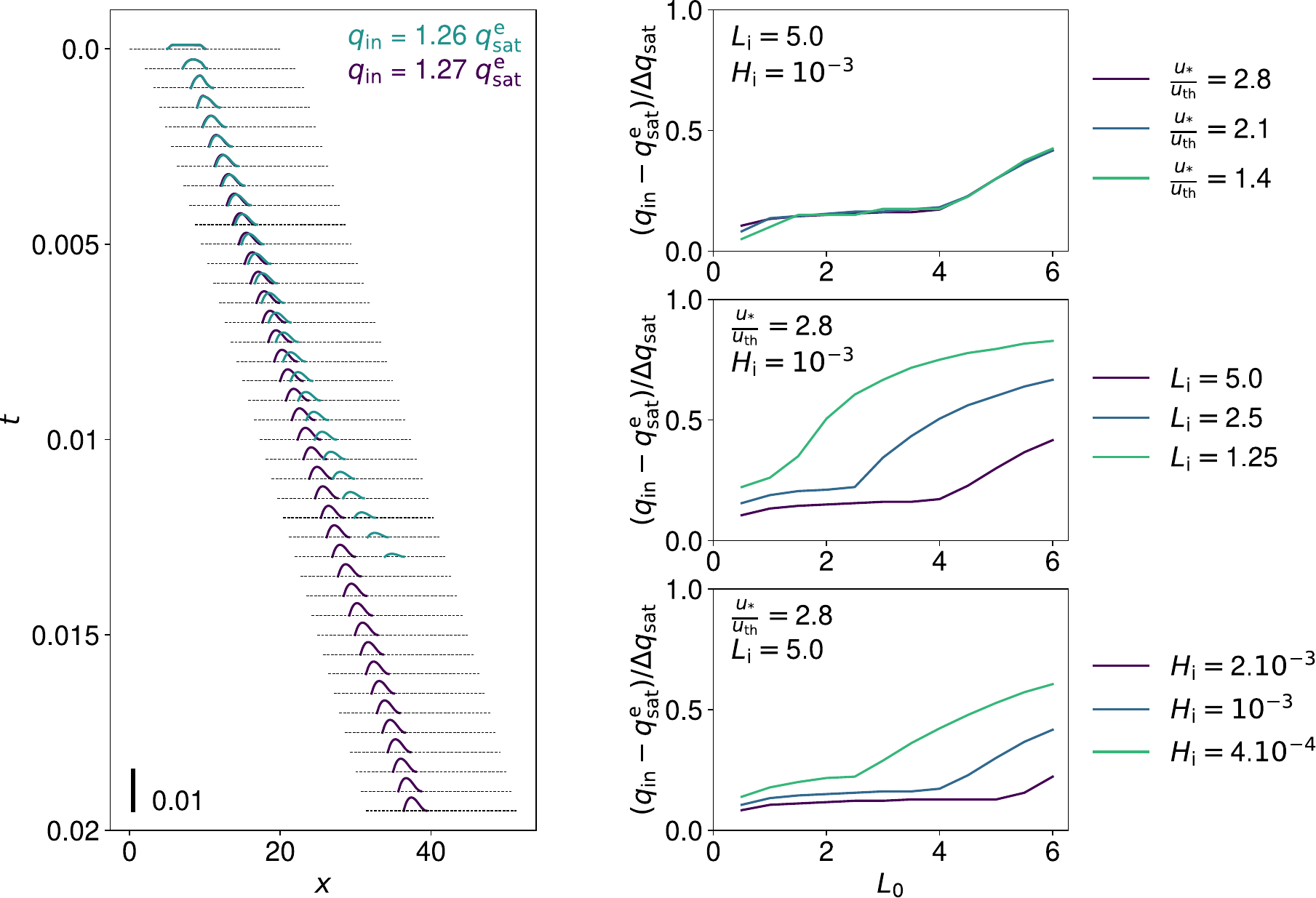}
\caption{
{\bf Influence of the initial conditions and wind speed on the limit between the growing and disappearing regimes.}
Left: Spatio-temporal diagram showing the evolution of the bedform just above and below the growth limit shown as a white line in Fig.
3 in main manuscript. These runs are for $L_0=2$, $u_*/u_{\rm th}=2.8$, $L_{\rm i}=5$ and $H_{\rm i}=10^{-3}$.
Right: Influence of wind speed, initial length and height (see values in legends) on this growth limit. Top: $L_{\rm i}=5$ and $H_{\rm i}=10^{-3}$. Middle: $u_*/u_{\rm th}=2.8$ and $H_{\rm i}=10^{-3}$. Bottom: $u_*/u_{\rm th}=2.8$ and $L_{\rm i}=5$. As shown in the top right panel, any dependence on wind velocity is accounted for once expressed in fluxes. The curves in the middle and bottom right panels show a first flat regime, independent of $L_0$, and then an increase of the input flux necessary to get the bedform growing. We find that the $L_0$-value of the transition depends on the initial mass of the bedform: it is larger for a larger mass. In other words, the dynamics starts to be sensitive to $L_0$ when the initial bedform mass is small, which requires a significantly larger input flux to grow.
}
\label{Supplfig:LimitDiagram}
\end{figure*}

\begin{figure*}
\centering
\includegraphics[width=\linewidth]{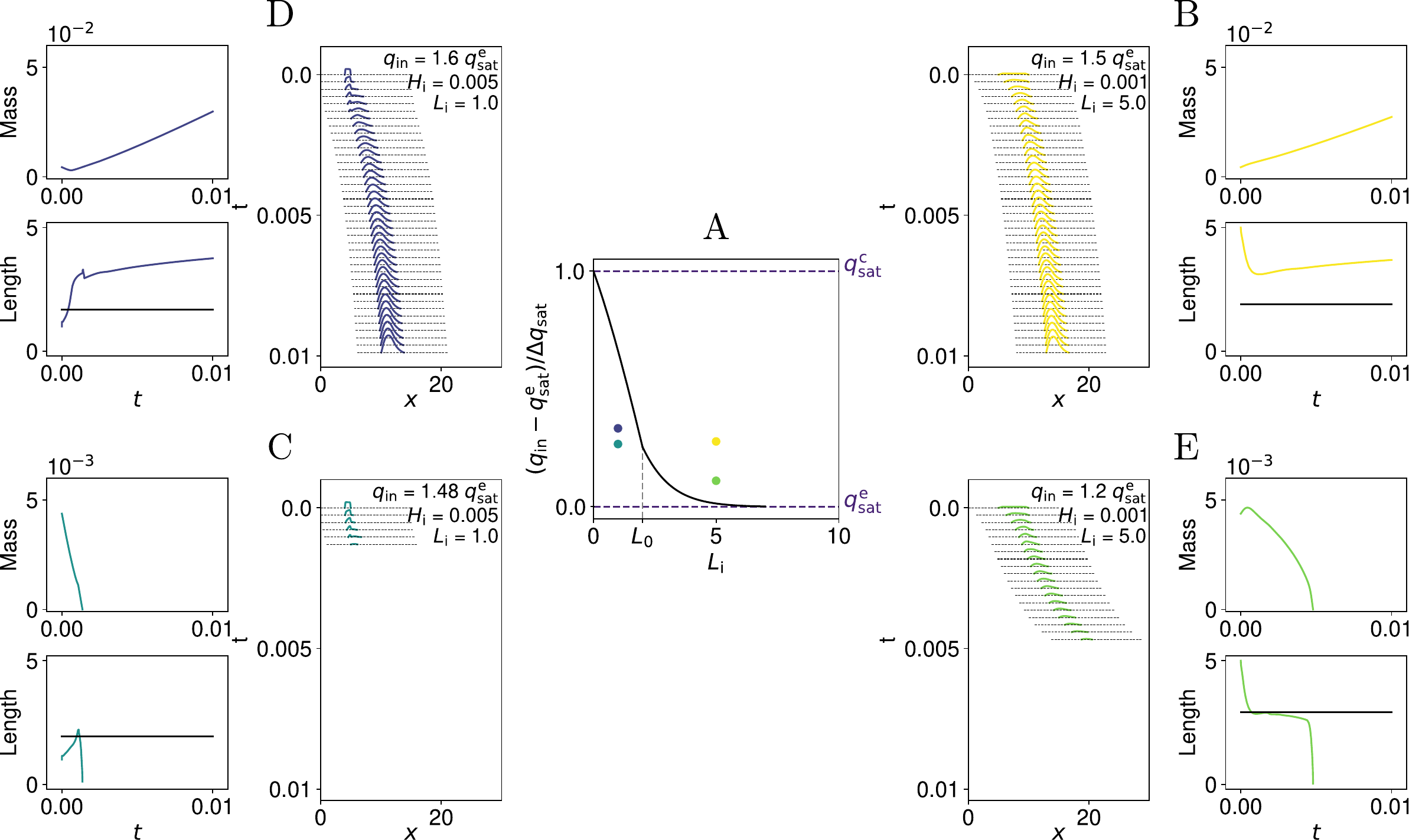}
\caption{
{\bf Comparison of simulations near the limit between the growing and disappearing regimes with the analytical solutions for flat case.}
Central panel {\bf A}: Parametric plane showing patch initial length $L_{\rm i}$ \emph{versus} input flux (normalized as $(q_{\rm in} - q_{\rm sat}^{\rm e})/\Delta q_{\rm sat}$, where $\Delta q_{\rm sat}=q_{\rm sat}^{\rm c} - q_{\rm sat}^{\rm e}$) to show four different cases ({\bf B}-{\bf E}, see corresponding colors and parameter values in legend). The black line represents Eqs.~\ref{qinvsLimassbalance}, and \ref{qinvsLimassbalancebis} of SI text associated with a mass-balanced flat patch ($q_{\rm in} = q_{\rm out}$). On the left of this curve, mass balance is negative ($q_{\rm in} < q_{\rm out}$), whereas it is positive on the right ($q_{\rm in} > q_{\rm out}$). In practice, the bedforms have a finite thickness, and this critical line is an approximation of a wider transition zone.
Lateral panels ({\bf B}-{\bf E}): spatio-temporal diagrams of these four cases, and time evolution of the bedform mass and length. In the panels for the length, the solid black line also represents the analytical limit $q_{\rm in} = q_{\rm out}$. All runs are for $L_0 = 2.0$.
A typical growing and migrating bedform is shown in case {\bf B}: it starts on the right of the mass-balanced curve, and gains mass with time. Meanwhile, its length, sufficiently larger than $L_0$, reduces as expected from the analytics (see also Fig.~\ref{Supplfig:FluxProfilesAnalytics}{\bf A}), and then gently increases, always staying well above the critical size (black line). A typical disappearing bedform is shown in case {\bf C}: it starts on the left of the mass-balanced curve, and looses mass with time. Its small initial length first increases, as expected from the analytics (see also Fig.~\ref{Supplfig:FluxProfilesAnalytics}{\bf D}), but its initial mass and the input flux were insufficient to make it cross the critical line significantly. The bedform then eventually dies. By contrast, case {\bf D} illustrates the example where the input flux was larger, and sufficient to allow the bedform to increase its length well over the critical size. As a result, even if its mass was initially decreasing, it eventually accumulates sand and grows. Finally, in case {\bf E}, the bedform initially gains mass while reducing its length as in {\bf B}, but the input flux was not sufficient to prevent it to cross the critical size.
}
\label{Supplfig:Li}
\end{figure*}

\begin{figure*}
\centering
\includegraphics[width=\linewidth]{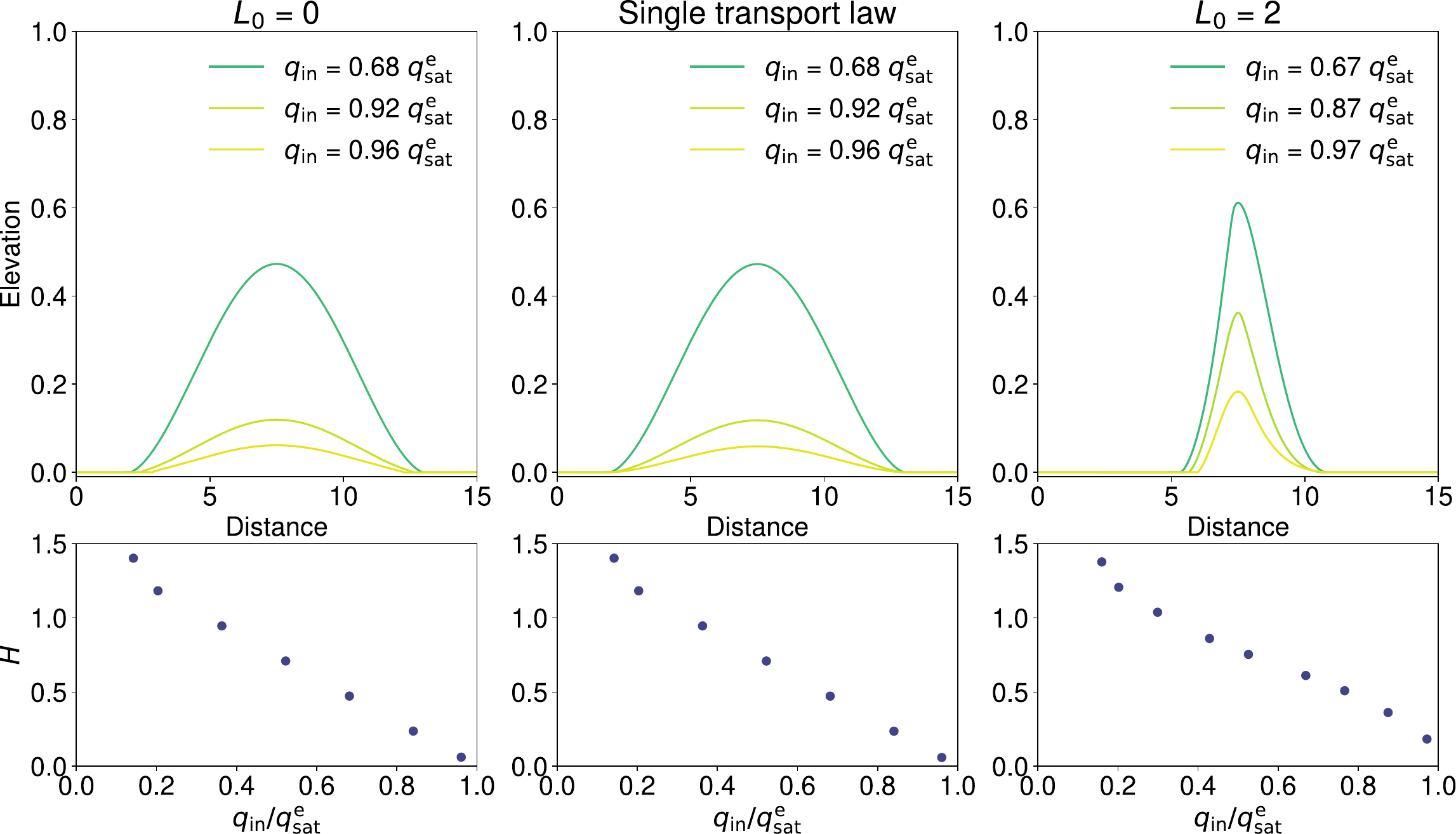}
\caption{
{\bf Steady propagative solutions at low incoming flux}
Stationary states (i.e. with the condition $q_{\rm in} = q_{\rm out}$) of the equations exist at low input flux: $q_{\rm in} < q_{\rm sat}^{\rm e}$. Top: corresponding elevation profiles $h(x)$ at different values of $q_{\rm in}$ (see legends) in the case of $L_0 = 0$ (left), for a single transport law (Eq.
4 of main manuscript) independent on the bed nature (middle), and for $L_0 = 2$ (right). Note the asymmetric profiles and reduced length in this later case, in comparison to the first two. Bottom: Decreasing crest height $H$ of these profiles as a function of $q_{\rm in}/q_{\rm sat}^{\rm e}$.
The profiles in the left and middle panels are identical as, for these low flux input conditions, the possibility of having a flux $q(x)$ larger than $q_{\rm sat}^{\rm e}$ never occurs when $L_0=0$.
}
\label{Supplfig:SteadyPropagativePatches}
\end{figure*}

\begin{figure*}
\centering
\includegraphics[width=0.85\linewidth]{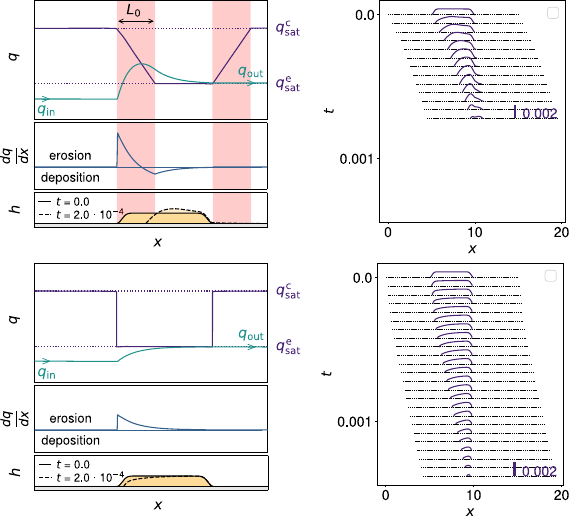}
\caption{
{\bf Dynamics of the bedform at low incoming sand flux}.
Schematic profiles and spatio-temporal diagrams as in Fig.
2 of main manuscript for $L_0=2$ (top) and $L_0=0$ (bottom).
In both cases, the bedform shrinks, as $q_{\rm out} \simeq q_{\rm sat}^{\rm e}$ is larger than $q_{\rm in }$. Their dynamics is not very different, except for a slightly enhanced upwind erosion for a finite $L_0$.
}
\label{Supplfig:lowqin}
\end{figure*}

\begin{figure*}
\centering
\includegraphics[width=0.75\linewidth]{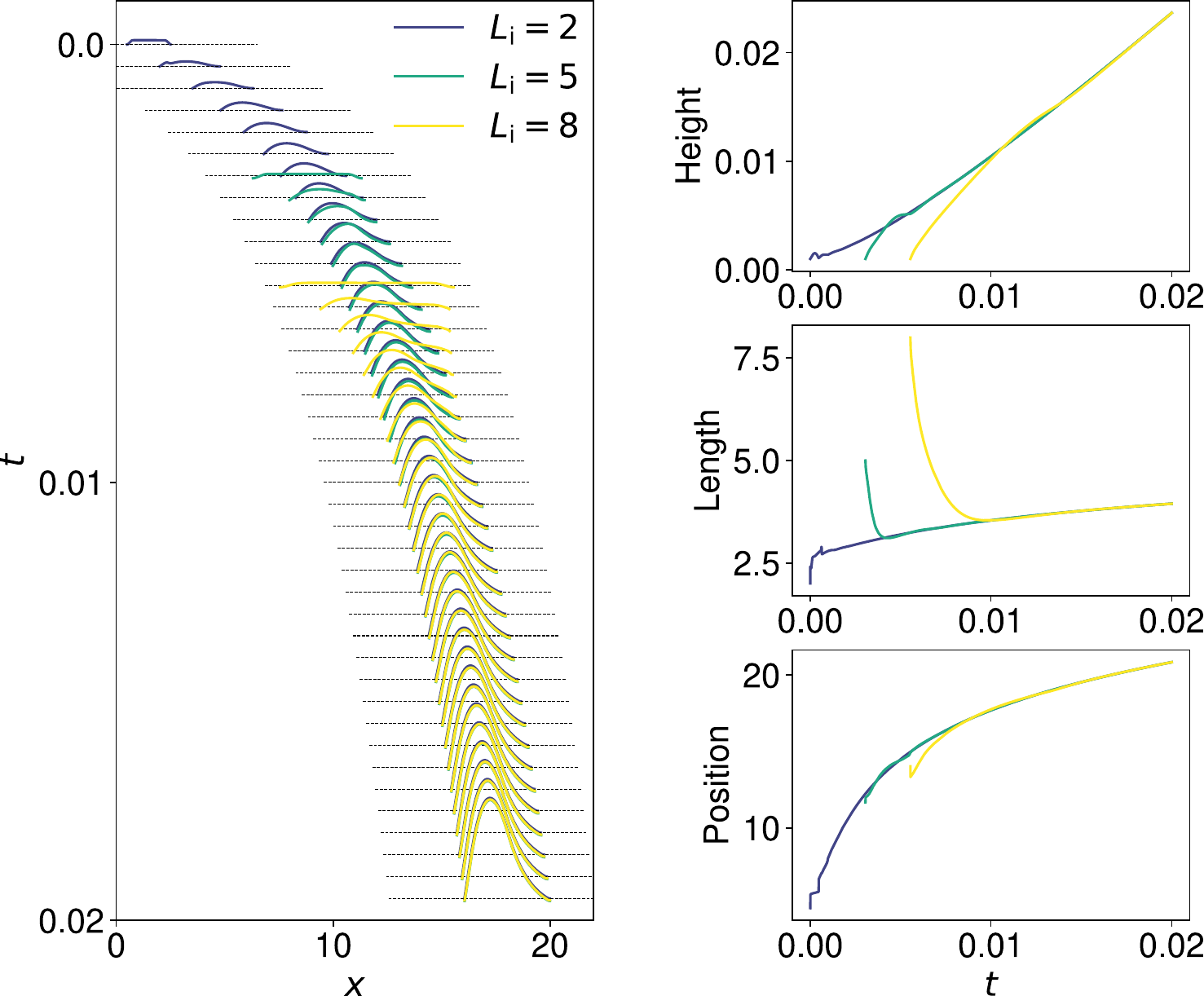}
\caption{
{\bf Influence of the initial length $L_{\rm i}$ on the evolution of the bedform}.
Spatio-temporal diagram (left) and time evolution of the bedform height (top), length (middle) and position (bottom) for three values of $L_{\rm i}= \{2,5,8\}$ . All these runs are for $L_0 = 2$ and $q_{\rm in} = 1.5\,q_{\rm sat}^{\rm e}$. Despite significant differences in shape over short times, the morphodynamics of the bedform eventually becomes independent of $L_{\rm i}$.
}
\label{Supplfig:InitialLength}
\end{figure*}

\begin{figure*}
\centering
\includegraphics[width=0.75\linewidth]{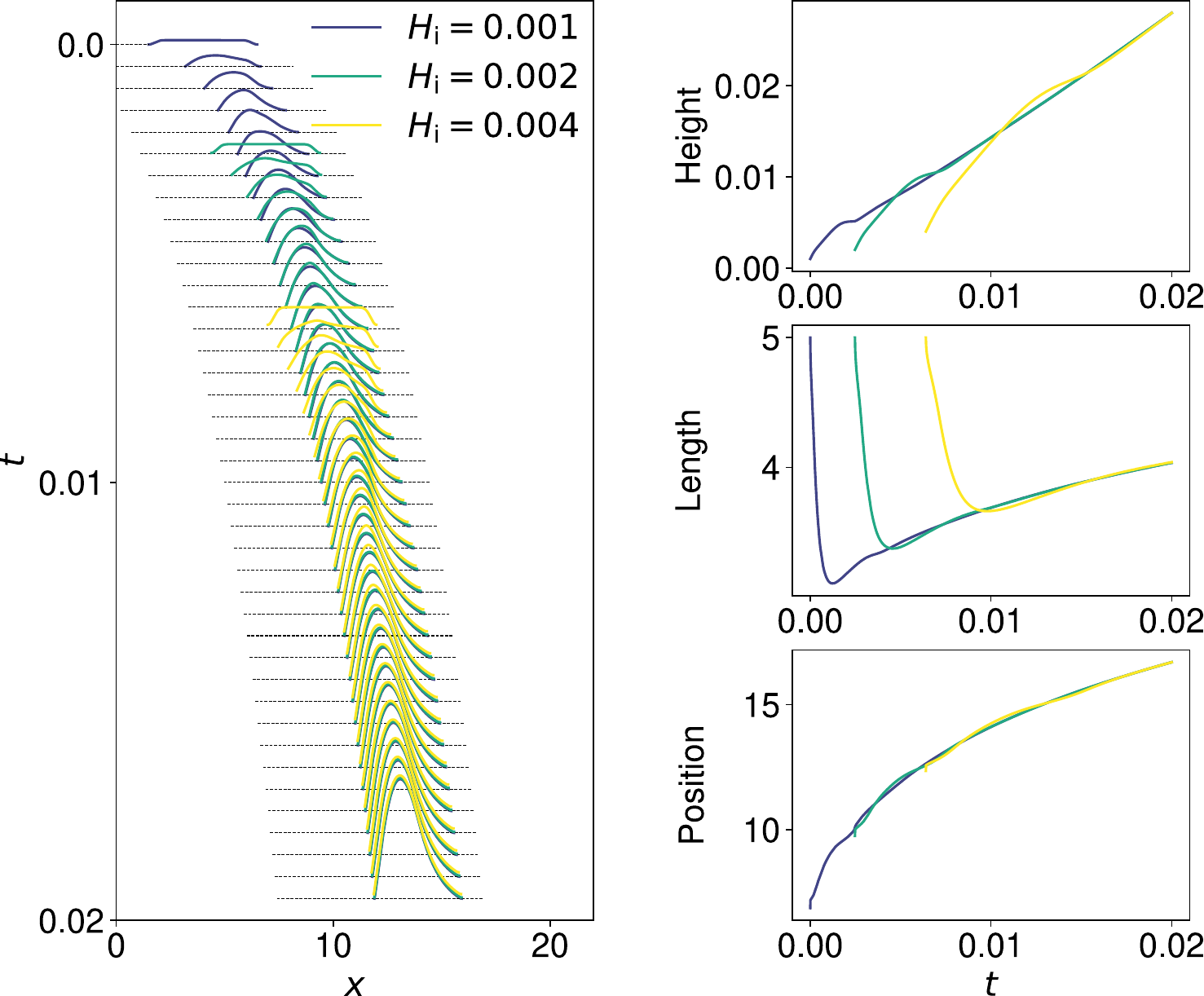}
\caption{
{\bf Influence of the initial height $H_{\rm i}$ on the evolution of the bedform}.
Spatio-temporal diagram (left) and time evolution of the bedform height (top), length (middle) and position (bottom) for three values of $H_{\rm i}= \{0.001,0.002,0.004\}$. All these runs are for $L_0 = 2$, $q_{\rm in} = 1.5\,q_{\rm sat}^{\rm e}$. Despite significant differences in shape over short times, the morphodynamics of the bedform eventually becomes independent of $H_{\rm i}$.
}
\label{Supplfig:InitialHeight}
\end{figure*}

\begin{figure*}
\centering
\includegraphics[width=0.85\linewidth]{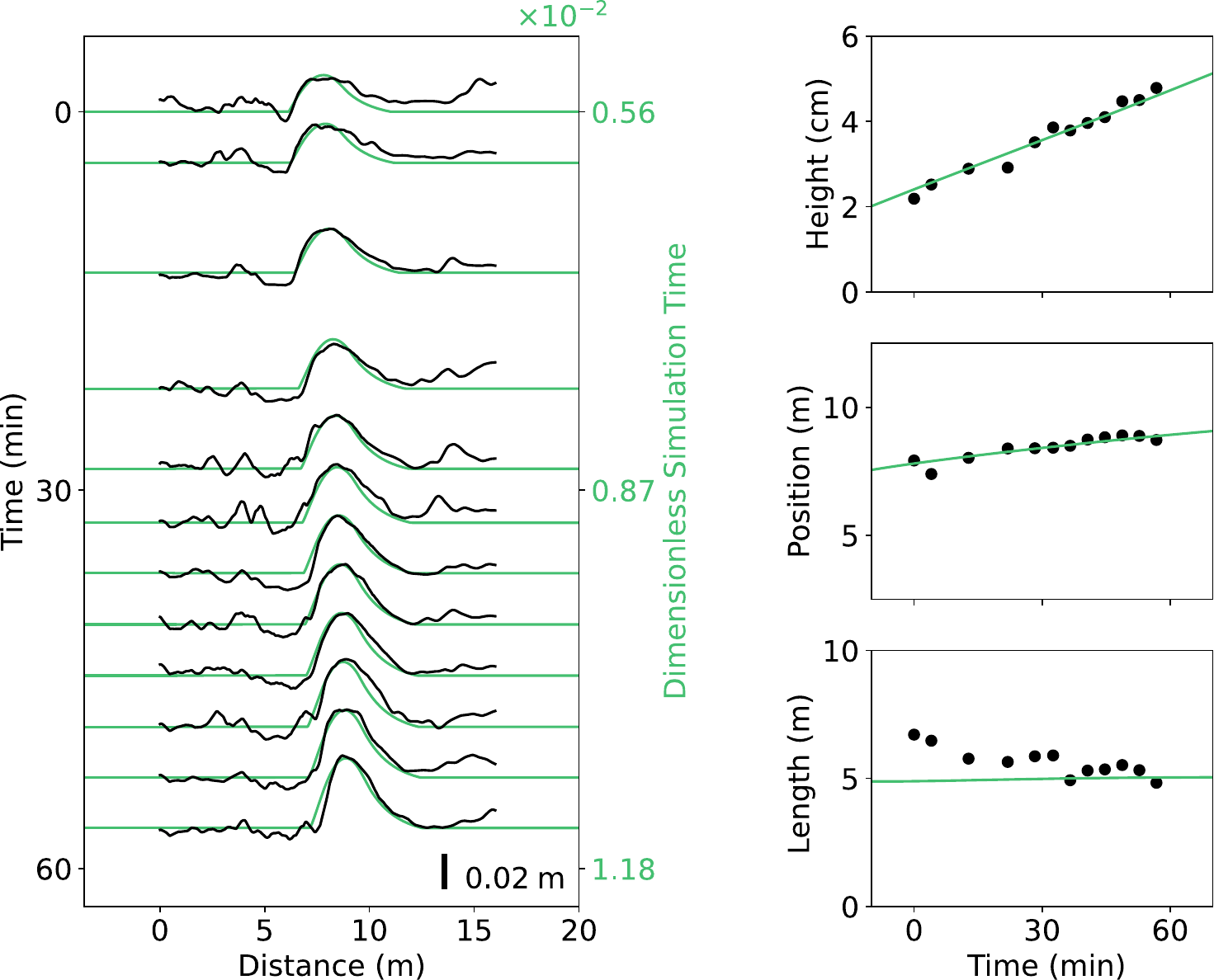}
\caption{
{\bf Model adjustment to field data showing the growth and migration of a meter-scale bedform.}
Same as Fig.
4 of main manuscript for another bedform nearby (same constant wind). Time origin is 10:04 am on the 13$^{\rm th}$ September 2022. The green solid lines are from the model with parameter values $\{L_0/L_{\rm sat},\, q_{\rm in}/q_{\rm sat}^{\rm e}\}=\{2.0,\,2.0 \}$. The ensemble of parameter values giving a reasonable fit of the data are displayed by the pink ellipse in Fig.~3 of the main manuscript. Right axis in green: dimensionless time in the simulation.
}
\label{Supplfig:SecondDataSetGrowthPropagation}
\end{figure*}

\begin{figure*}
\centering
\includegraphics[width=\linewidth]{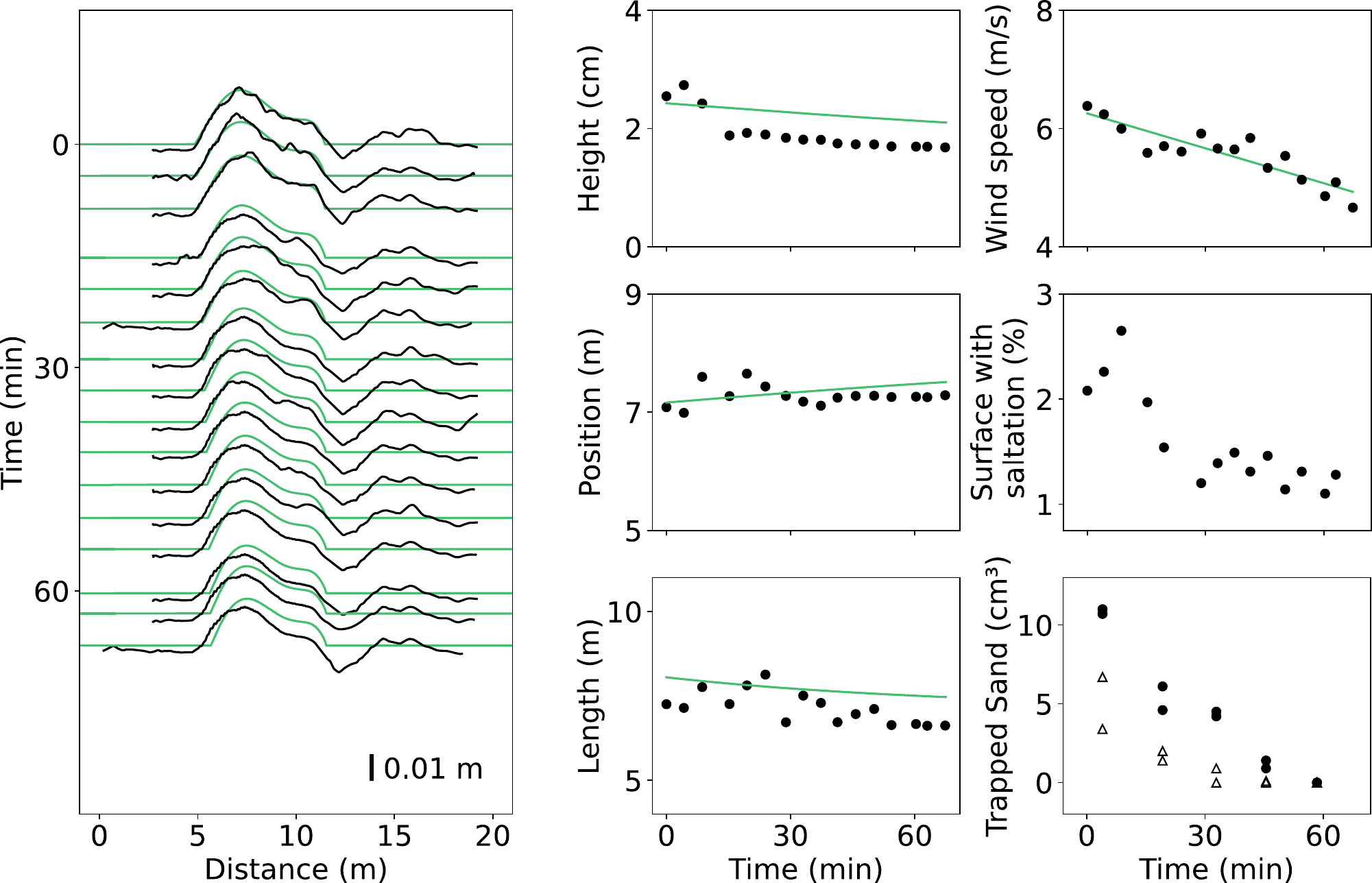}
\caption{
{\bf Model adjustment to field data showing a disappearing meter-scale bedform.}
Same as Fig.
4 of main manuscript and Fig. \ref{Supplfig:SecondDataSetGrowthPropagation}, but here for the case of a decreasing wind. Time origin is 10:58 am on the 12$^{\rm th}$ September 2023. The green solid lines are from the model with best fit parameter values $\{L_0/L_{\rm sat} ,\, q_{\rm in}/q_{\rm sat}^{\rm e}\}=\{0.0,\,0.5 \}$. Note that as $q_{\rm sat}^{\rm e}$ depends on the wind shear velocity (Eq.~4 of main manuscript), the input flux $q_{\rm in}$ decreases in proportion. A finer adjustment would consist of changing $q_{\rm in}$ at a different rate than $q_{\rm sat}^{\rm e}$. Besides, panels in the right column display time variations of wind speed measured at $0.24$~m height, the percentage of the surface where saltation was detected during TLS scan, and sand captured in traps downwind ({\large $\bullet$}) and sideways ($\triangle$) of the bedform. This provides evidence of sediment transport during this period, decreasing and almost vanishing after one hour. Comparing data from the two sand traps, there is a larger flux out of the bedform than on its side, which gives a proxy for the flux upwind of the bedform. This is consistent with a shrinking dynamics.
}
\label{Supplfig:FitDecroissance}
\end{figure*}

\begin{figure*}
\centering
\includegraphics[width=0.6\linewidth]{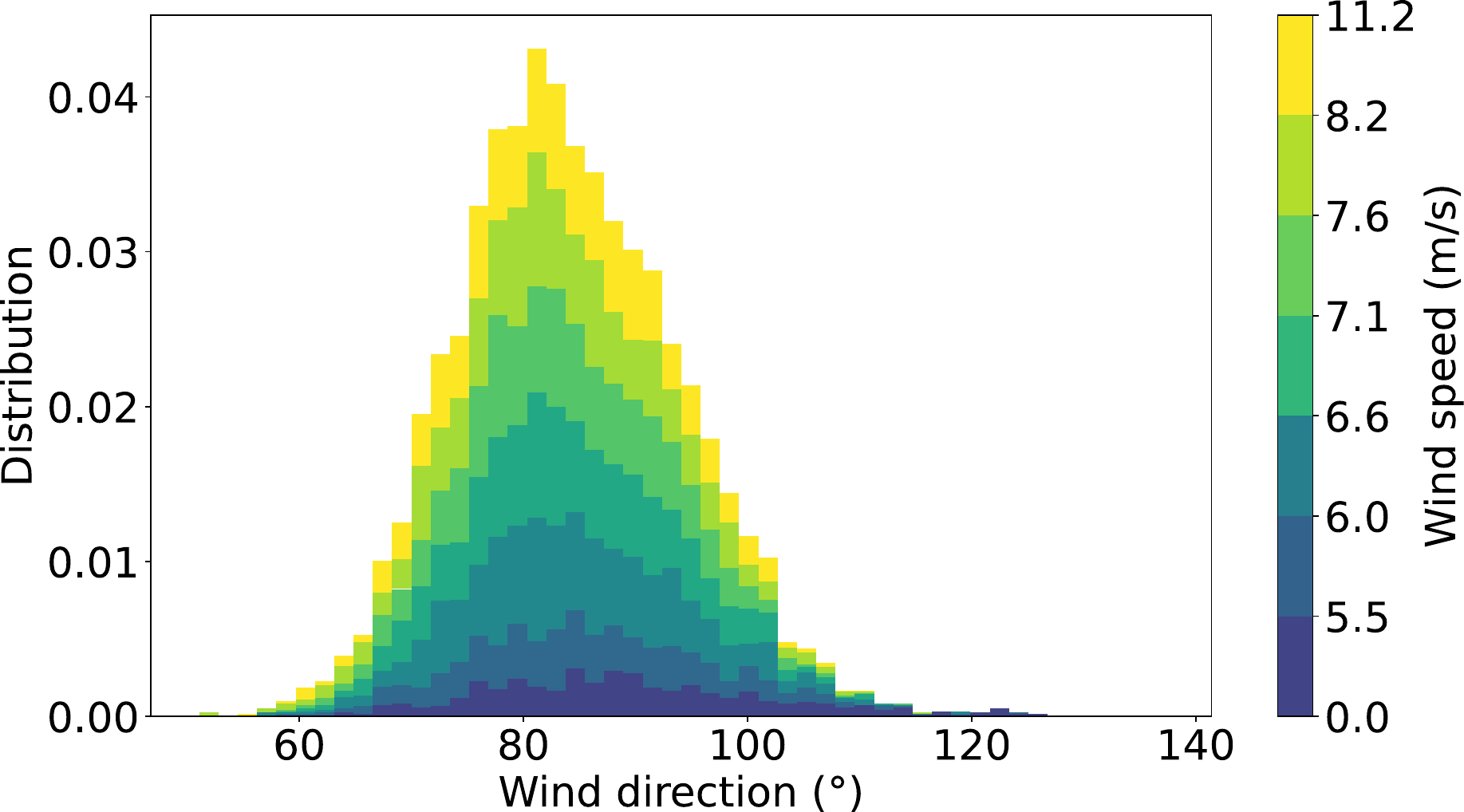}
\caption{
{\bf Wind orientation distribution}. Wind speed and orientation corresponding to the measurements on the 13$^{\rm th}$ September 2022 (Fig.
1B,C). Data cumulated over the whole period of profile acquisition. Wind orientation is given in degrees anticlockwise from North. This peaked distribution corresponds to a unimodal wind regime with a mean orientation of $84^\circ$ (easterly wind) with an angular dispersion of $7^\circ$.
}
\label{Supplfig:WindOrientationDistribution}
\end{figure*}

\begin{figure*}
\centering
\includegraphics[width=\linewidth]{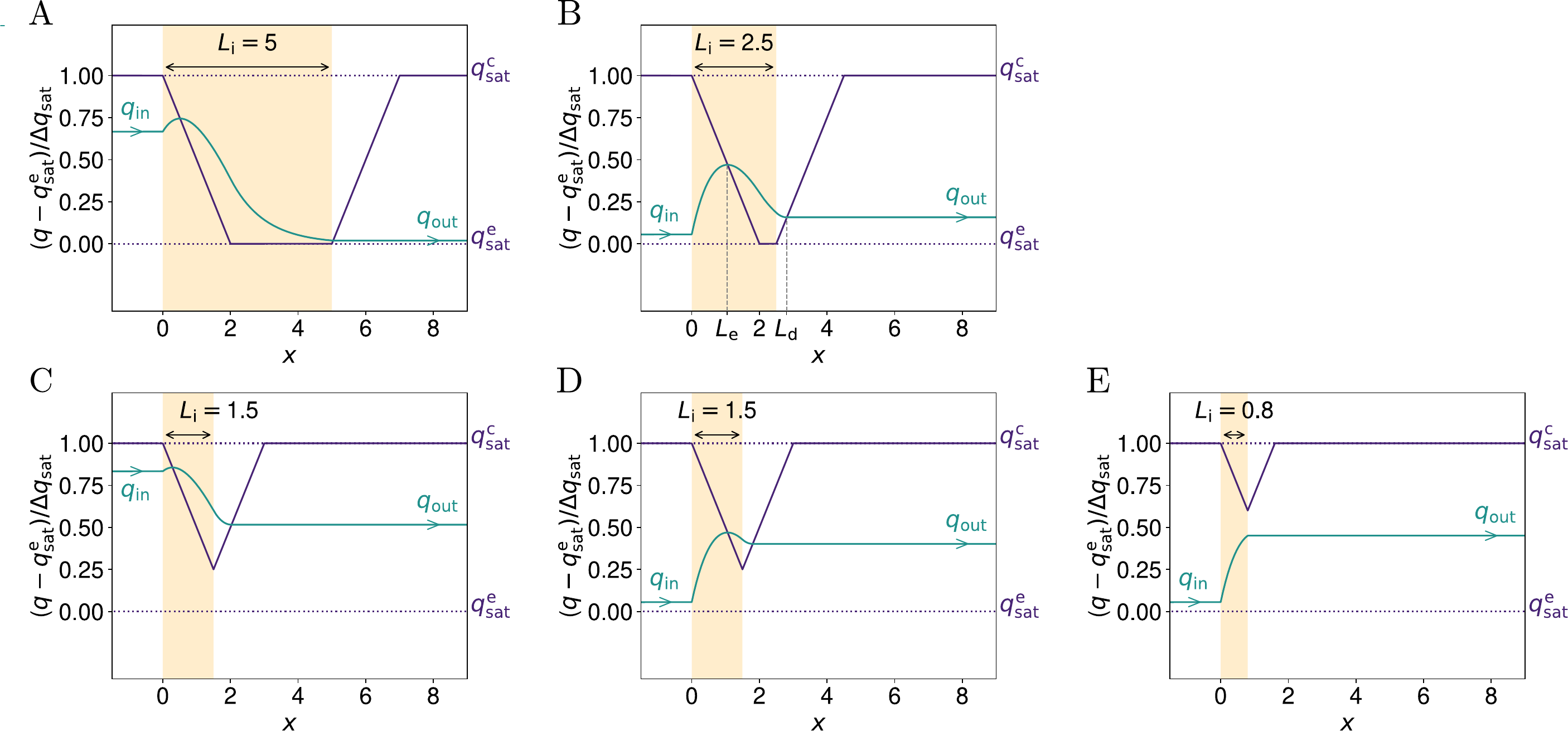}
\caption{
{\bf Sand flux profiles from analytical calculations associated with perfectly flat beds}.
Profiles computed from expressions in SI text, with $L_0 = 2$, $q_{\rm sat}^{\rm c}=2.8q_{\rm sat}^{\rm e}$, and various patch lengths $L_{\rm i}$ (see values in legend) and input fluxes:
({\bf A}) 
$q_{\rm in} = 2.2q_{\rm sat}^{\rm e}$ and long initial patch length $L_{\rm i}>L_0$.
In this case, $q_{\rm in} > q_{\rm out}$, as the output flux is close to $q_{\rm sat}^{\rm e}$. The zone eroded at the upwind edge (length $L_{\rm e}$) is larger than the deposition zone downwind of the bedform (length $L_{\rm d}-L_{\rm i}$). Such a patch will gain mass as well as reduce in length.
({\bf B}) 
$q_{\rm in} = 1.1q_{\rm sat}^{\rm e}$ and long initial patch length $L_{\rm i}>L_0$.
Here $q_{\rm in} < q_{\rm out}$. Such a patch will both lose mass and reduce in length. Dashed lines : display of the lengths $L_{\rm e}$ and $L_{\rm d}$.
({\bf C})
$q_{\rm in} = 2.5q_{\rm sat}^{\rm e}$ and short initial patch length $L_{\rm i}<L_0$ with $L_{\rm i}>L_{\rm e}$.
Here the patch will gain mass as $q_{\rm in} > q_{\rm out}$, but in contrast to case {\bf A}, the erosion zone upwind is smaller than the deposition zone downwind, and it will consequently increase in length.
({\bf D})
$q_{\rm in} = 1.1q_{\rm sat}^{\rm e}$ and short initial patch length $L_{\rm i}<L_0$ with $L_{\rm i}>L_{\rm e}$.
This case behaves similarly to {\bf B}.
({\bf E})
$q_{\rm in} = 1.1q_{\rm sat}^{\rm e}$ and short initial patch length $L_{\rm i}<L_0$ with $L_{\rm i}=L_{\rm e}$.
The whole patch is in the erosion zone, and will disappear.
}
\label{Supplfig:FluxProfilesAnalytics}
\end{figure*}

\begin{figure*}
\centering
\includegraphics[width=\linewidth]{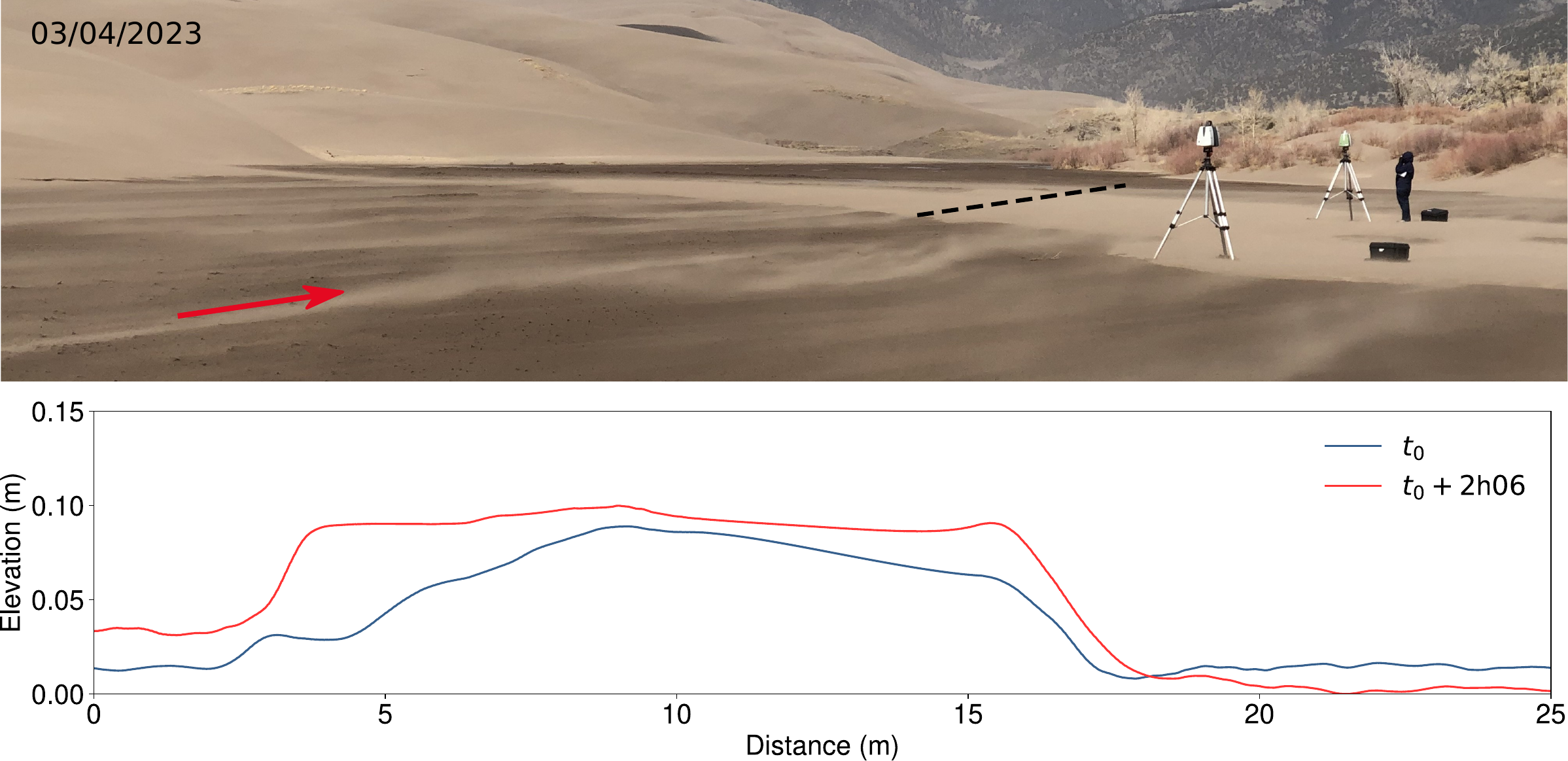}
\caption{
{\bf Spreading dynamics}.
Top : Photo of a spreading bedform at Great Sand Dunes National Park (USA) on the 3$^{\rm rd}$ April 2023. Here the consolidated surface is moist sand. The averaged wind speed $u=9.33$~m/s was measured $220$~m away from the bedform at $z=0.24$~m using a 3D Sonic anemometer. Red arrow indicates wind direction. Black dashed line shows the approximate position of displayed transects.  Bottom : Elevation profiles of the bedform measured over 2 hours and 06 minutes, starting at $t_0=$12:18. Sand accumulates at both upwind and downwind toes of the bedform, showing a spreading dynamics. Note the relatively large scale of this bedform, in comparison to those in the growing and migrating regime observed in Namibia (Fig.~4).
}
\label{Supplfig:Spreading}
\end{figure*}

\begin{figure*}
\centering
\includegraphics[width=\linewidth]{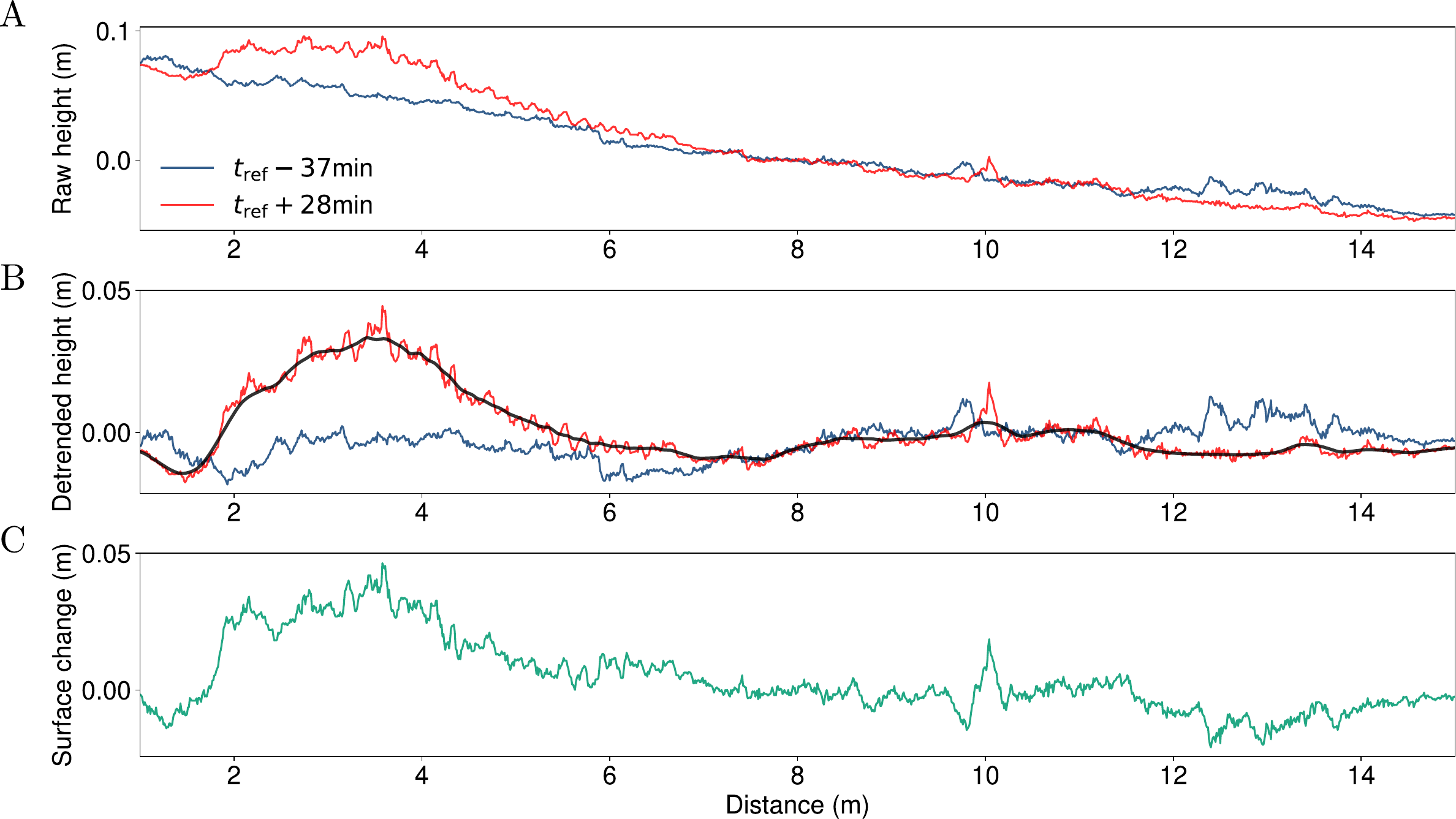}
\caption{
{\bf Detrending and smoothing data processes}.
Illustration of data processing to obtain elevation profiles of the bedform displayed in Fig.~4 of main manuscript. $t_{\rm ref}$ corresponds to 9:40 am on the 13$^{\rm th}$ September 2022.
({\bf A}) 
Elevation profiles as obtained by the TLS raw signal once filtered to remove saltation and gridded at $1$~cm horizontal resolution. Blue line: initial surface before bedform emergence (09:03 am). Red line: Bedform at time 10:08 am. 
({\bf B})
Detrending: the underlying topography is subtracted as explained in the section describing the field measurements method in the main manuscript. Smoothing: the black line is the result of a $45$~cm mean moving window filter on the red profile in order to remove the sand ripples. This detrended and smooth profile is displayed in Fig.~4 (8th profile) and has been fitted by the model.
({\bf C}) 
Surface difference where the blue profile is subtracted from the red one, in order to emphasize erosion and deposition with respect to the consolidated bed.
}
\label{Supplfig:Detrending}
\end{figure*}

\begin{figure*}
\centering
\includegraphics[width=0.8\linewidth]{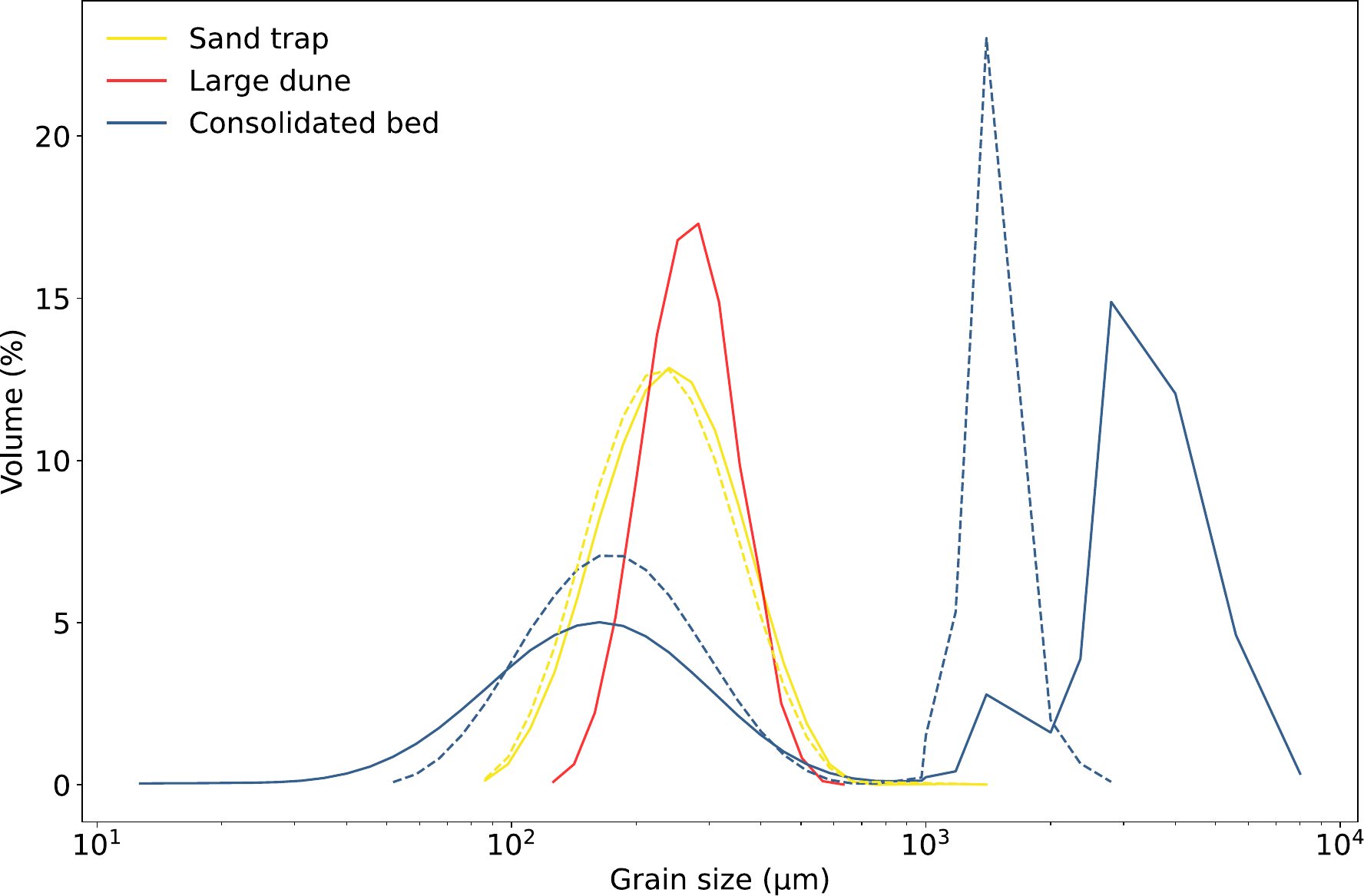}
\caption{
{\bf Grain size distributions}.
Size distributions of saltating particles collected in sand traps (yellow), on a nearby large dune located at 15$^\circ$01'35.14''E, 23$^\circ$34'35.48''S (red), and on the consolidated bed in the interdune area (blue). Sand traps: one was located next to a bedform (that displayed in Fig.~\ref{Supplfig:FitDecroissance}) but on the consolidated bed; the other one was deployed on a nearby flat sand bed. The close correspondence of the two yellow curves demonstrates that saltating grains are the same on both bed types. Large dune: the red distribution is similar to the yellow ones, which suggests that the transported grains come from the neighboring dune. Consolidated bed: the distributions are clearly bimodal, with a peak at small size similar to the sand bedforms, and peaks associated with millimeter-size gravels. These large particles typically represent $\simeq 40$\% of the bed material. The two blue curves correspond to two similar places between the bedforms, but show a spatial variation in gravel size.
}
\label{Supplfig:Grainsize}
\end{figure*}

Fig.~\ref{Supplfig:tauspreading} complements Fig.~2 of main manuscript by showing the shear stress profile.\\
Fig.~\ref{Supplfig:patchlife} complements Fig.~2 of main manuscript by showing the time evolution of the bedform's height, position and length.\\
Fig.~\ref{Supplfig:varyL0} displays the model's behavior for three different $L_0$-values and $q_{\rm in}=2q_{\rm sat}^{\rm e}$.\\
Fig.~\ref{Supplfig:varyqin} displays the model's behavior for three different $q_{\rm in}$-values and $L_0 = 2L_{\rm sat}$.\\
Fig.~\ref{Supplfig:L0zero} displays the model's behavior for three different $q_{\rm in}$-values and $L_0 = 0$.\\
Fig.~\ref{Supplfig:LimitDiagram} shows how the transition from the disappearing to the growing regimes is modified when the model's parameters are varied.\\
Fig.~\ref{Supplfig:Li} shows the comparison of simulations near the transition from the disappearing to the growing regimes with the analytical solution considering flat beds. \\
Fig.~\ref{Supplfig:SteadyPropagativePatches} illustrates the effect of the two transport laws and the transition length $L_0$ on the steady propagative solutions. \\
Fig.~\ref{Supplfig:lowqin} displays the model's behavior for low $q_{\rm in}$-values. \\
Fig.~\ref{Supplfig:InitialLength} shows how the bedform evolution is affected by the initial length $L_{\rm i}$. \\
Fig.~\ref{Supplfig:InitialHeight} shows how the bedform evolution is affected by the initial height $H_{\rm i}$. \\
Fig.~\ref{Supplfig:SecondDataSetGrowthPropagation} presents the comparison of the model with a growing and migrating bedform (see pink ellipse in Fig.~3 of main manuscript).\\
Fig.~\ref{Supplfig:FitDecroissance} presents the comparison of the model with a disappearing bedform.\\
Fig.~\ref{Supplfig:WindOrientationDistribution} complements Fig.~1 of main manuscript by showing the distribution of the wind direction during the experiment on the 13$^{\rm th}$ September 2022. \\
Fig.~\ref{Supplfig:FluxProfilesAnalytics} shows the analytical flux profiles calculated in this SI appendix. \\
Fig.~\ref{Supplfig:Spreading} displays an example of a spreading bedform. \\
Fig.~\ref{Supplfig:Detrending} presents the detrending and smoothing of the TLS data. \\
Fig.~\ref{Supplfig:Grainsize} shows grain size distributions from various locations.

\newpage

\newpage

\newpage

\newpage

\newpage

\newpage

\newpage

\newpage

\newpage

\newpage
\newpage

\newpage

\newpage

\newpage

\newpage

\newpage

\newpage

\newpage

\clearpage
\newpage
\bibliographystyle{elsarticle-num}
\bibliography{dune}

\end{document}